\documentclass[fleqn,usenatbib]{mnras}

\usepackage{newtxtext,newtxmath}
\usepackage{xcolor}

\usepackage[T1]{fontenc}

\DeclareRobustCommand{\VAN}[3]{#2}
\let\VANthebibliography\thebibliography
\def\thebibliography{\DeclareRobustCommand{\VAN}[3]{##3}\VANthebibliography}

\usepackage{graphicx}	
\usepackage{amsmath}	

\usepackage{hyperref}
\DeclareRobustCommand{\citelinktext}[2]{%
	\hyperlink{cite.#1}{#2}%
}

\title[Clumps in spiral galaxies at $z \lesssim 3$]{Clumps in spiral galaxies at $z \lesssim 3$: Disentangling two spatial modes of star formation}

\author[I. V. Chugunov, A. A. Marchuk]{
Ilia V. Chugunov$^{1, 2}$\thanks{E-mail: 21ilya07@gmail.com}
and Alexander A. Marchuk$^{1, 3}$
\\
$^{1}$Pulkovo Astronomical Observatory, Russian Academy of Sciences, Pulkovskoye Chaussee 65/1, 196140 St. Petersburg, Russia\\
$^{2}$Sternberg Astronomical Institute, Lomonosov Moscow State University, Universitetsky Pr. 13, 119234 Moscow, Russia\\
$^{3}$Saint Petersburg State University, 7/9 Universitetskaya Nab., St. Petersburg 199034, Russia\\
}

\date{Accepted XXX. Received YYY; in original form ZZZ}

\pubyear{2026}

\begin{document}
\label{firstpage}
\pagerange{\pageref{firstpage}--\pageref{lastpage}}
\maketitle

\begin{abstract}
At high redshifts, star formation in galaxies is more often concentrated in clumps than in spiral arms. Although clumps are actively studied, it is rarely done considering spiral arms as objects for study as well, and the connection between clumps and spirals remains understudied. We used a sample of 159 spiral galaxies at $0.1 \leq z \leq 3.3$ observed by HST and JWST. Using the residual images from photometric models with spiral arms constructed before, which was not done previously, we have done identification of clumps and measured their properties with photometric decomposition, finding 3003 clumps in overall, and performing SED fitting for a fraction of them. We examined the overall properties of clumps, focusing on the properties of spiral structure. We found that clumps luminosities, masses and sizes are smaller than commonly accepted in literature, either for the reason spiral arms were modelled separately, or because clumps in spiral galaxies are different compared to clumpy ones. We demonstrate the connection between clumps and spirals, in particular clumps are spatially concentrated towards spirals and various parameters of clumps and spirals correlate. Also, clumps in spiral arms tend to be smaller but brighter compared to clumps in inter-arm area, but their colours are similar. There are also some differences between clumps and spirals such as colour, emphasizing the importance of their separate analysis. Our results probably indicate that spiral arms stimulate the formation of clumps, although star formation properties of clumps in spiral arms is not changed compared to inter-arm region.
\end{abstract}




\section{Introduction}
\label{sec:introduction}
In the local Universe, star formation on galactic scales is mostly structured in a more or less defined spiral pattern. There is no wonder that the majority of bright galaxies (75\% objects with absolute magnitude $M(B) < -20$, or with stellar mass $M_* > 10^{10} M_{\sun}$) are spirals \citep{Conselice2006, Willett2013, Kelvin2014}. It is well-known that spirals are more then just regions of increased star formation appearing brighter than background disc. In the majority of cases, these structures also concentrate gas and stellar mass. Therefore, spiral arms can influence their host galaxy in a variety of ways. They driving its secular evolution by angular momentum redistribution, disc dynamical heating and other effects \citep{Sellwood2014}. However, the mechanisms responsible for the formation of spiral structure in galaxies are still debated \citep[see reviews by][]{Dobbs2014, Sellwood2022a}.

The fraction of spiral galaxies depends significantly on the mass and environment. First, spiral galaxies become less abundant at lower and higher mass ends. Specifically, in $M_* < 10^{9.5} M_{\sun}$ mass range star-forming galaxies are predominantly irregular, lacking a defined spiral pattern, and ellipticals constitute the majority at $M_* > 10^{11} M_{\sun}$ \citep{Kelvin2014}. Also, spirals are much less common in dense environments: spiral galaxies contribute less than 10\% to total number density in the central regions of clusters, but 80\% in the field \citep{Dressler1980, Houghton2015}. These trends show that the formation of spiral structure and, in broader perspective, galactic star formation are regulated by the conditions and characteristics of a galaxy.

In the same time, the ubiquity of spiral structure is not the case at high redshifts. Since the start of its mission in 1990, the Hubble Space Telescope (HST) has revealed the rest-frame UV structure of high-$z$ galaxies with great detail. Soon after, it became evident that distant galaxies look much more clumpy and asymmetric than their counterparts from the local Universe \citep{vandenBergh1996}. Many star-forming galaxies at $z > 1$ lack any spiral structure at all and consist of multiple so-called clumps \citep[see, e.g. a review by][]{Conselice2014}. The fraction of clumpy galaxies is well established to have a peak of nearly $\sim 60\%$ at $z \sim 2$ \citep[e.g.][]{Murata2014, Shibuya2016, Sattari2023, delaVega2025, Sok2025}, which corresponds to the peak of star formation in the Universe, also known as ``cosmic noon'' \citep{Madau2014}.

Nevertheless, spiral galaxies still exist at high $z$, although they become less common with increasing redshift. For instance, their intrinsic fraction is believed to be about 10--40\% near $z \sim 2$ \citep{Margalef-Bentabol2022, Kuhn2024, EspejoSalcedo2025}. Observational effects make it difficult to identify spiral galaxies reliably, and actual observed fractions of spirals at such redshifts are around 10\% or even less. Nowadays, there are a few individual examples of known spiral galaxies at $z > 4$, with highest $z = 5.2$ \citep{Tsukui2021, Xiao2025, Ikeda2025, Jain2025}. In the same time, clumpy galaxies exist not only at $z > 1$: with the help of large surveys, examples of clumpy galaxies in the local Universe were discovered \citep[e.g.][]{Adams2022, Santos-Junior2025}. There is also a class of so-called ``clumpy spirals'' which are galaxies where clumps and spiral structure coexist; possibly, they can represent a transitional stage between true clumpy galaxies and regular spirals \citep{Elmegreen2005, Elmegreen2007}.

What are clumps themselves? There is still no formal definition which observed structures should be called so, but usually this term implies kiloparsec or sub-kiloparsec scale features that are visible in the rest-frame blue and ultraviolet parts of the spectrum. This indicates that clumps are dominated by young stellar population and characterised by active star formation \citep{Guo2015}. For an individual galaxy, clumps' combined contribution to its total SFR can reach values up to 30\% \citep{Kalita2025a}.

On the other hand, it was long unclear if clumps host any significant stellar mass overdensity, or they are simply star-forming regions \citep[see, for example][]{Buck2017}. This changed with the launch of James Webb Space Telescope (JWST) in 2021 which made rest-frame infrared imaging of distant galaxies available. It turned out that at least some clumps are also well observed in this part of the spectrum and allowed to measure their masses robustly. Measurements have shown that clumps' mass upper limit can be as high as $10^{9.5} M_{\sun}$ or even approach tremendous $10^{10} M_{\sun}$. Such large values should raise questions regarding clumps' influence on the dynamics of the host galaxy. The detection limit is typically $10^{7\ldots8} M_{\sun}$ at $z \sim 1$ \citep{Kalita2025a, Kalita2025b, Sok2025}. For an individual galaxy, clumps can constitute up to 20\% of its total mass \citep{Kalita2025a}. Of course, in pre-JWST era efforts were also made to measure clump masses, but without rest-frame infrared data their results were inconclusive and often opposed one another. For example, \citet{Soto2017} reported clump contributions to the total galaxy mass as high as 93\%, whereas \citet{Wuyts2012} measured highest values of 7\%, significantly smaller than their observed SFR fraction.

Regarding their sizes, clumps can have half-light radii ($r_e$) approaching 2--3 kpc. However, most of them are on the order of hundreds of parsecs in size, which is already at the resolution limit for JWST for $z > 1$ \citep{Kalita2025a}. Noteworthy, there are some studies of strongly lensed distant clumpy galaxies, where the resolution limit is pushed down to a hundred or even ten of parsecs, and clumps as small as these are also observed \citep{Livermore2015, Dessauges-Zavadsky2017, Claeyssens2023, Claeyssens2025}.

Another extensively studied aspects of clumps are their kinematics and gas content. These are primarily obtained by spectroscopic studies using largest optical telescopes, such as Very Large Telescope (VLT), and, with a slightly different scope, with Atacama Large (Sub-)Millimetre Array (ALMA) \citep{Epinat2011, Hodge2020}. Kinematic studies of clumps indicate that most of them are genuine disc objects, but not all of them \citep{Puech2010, Genzel2011}. This leads to a distinction between \textit{in situ} and \textit{ex situ} clumps: the former have disc kinematics and presumably formed in a galaxy, whereas the latter are external objects with typically higher ages and masses \citep{Mandelker2017}. Regarding gas in clumps, it was found with ALMA that some of UV-observed clumps do not show any overdensities in cold gas \citep{Hodge2016, Cibinel2017}, and some do, possibly with offsets \citep{Carniani2017,Dessauges-Zavadsky2023}. These distinctions and the overall diversity of their parameters makes ``clumps'' rather overloaded term, defined mostly from the observational perspective.

With all these properties combined, clumps resemble ordinary star-forming regions in the local Universe, but their sizes and masses are significantly larger. This difference in sizes can be explained by different conditions in discs of high-$z$ and local galaxies. Namely, discs of distant galaxies have higher velocity dispersion \citep{Genzel2006, Law2009}, higher gas fraction and smaller disc size, so their gas surface density is also larger \citep{Carilli2013}. This leads to both larger Jeans instability length and critical length of perturbations which are stabilized by differential rotation which serve as the lower and the higher size limits of gravitationally collapsing regions, respectively \citep[see Chapter 6.2 in][]{Binney2008}. The disc itself undergoes a violent disc instability (VDI) state where marginally unstable state is self-sustained (see \citealt{Dekel2009, Cacciato2012, Forbes2014} for the theoretical framework, and \citealt{Safronov1960, Toomre1964, Goldreich1965a, Jog1984} for original formulation of instability criteria). Simulations also support this scenario \citep[e.g.][]{Tamburello2015, Mandelker2017}.

There is a matter of dispute regarding the lifespan and survival of clumps. Some studies and simulations claim that \textit{in situ} clumps (at least massive ones) are long-lived and can survive up to a few hundred Myr \citep{Mandelker2014, Mandelker2017}, while others argue that they are short-lived and will be destroyed by stellar feedback in about tens of Myr \citep{Hopkins2012, Buck2017, Oklopcic2017, Meng2020}. Note that \textit{ex situ} clumps are expected to have ages of at least $\gtrsim 1$ Gyr \citep{Mandelker2017, Kalita2025a}. From the observational perspective, efforts to measure clump ages are made, but neither long-lived nor short-lived scenario was not ruled out decisively \citep{Kalita2025a, Sok2025}. In turn, the question of clump lifespans have significant consequences. There are both theoretical \citep{Dekel2009} and observational \citep{Genzel2011} considerations supporting clump inwards radial migration, although some studies do not find any evidence of migration, or state that only a minority of clumps undergoes it \citep{Mandelker2017, Oklopcic2017}. If clumps are long-lived, then migration is expected to end with clump infalling into bulge or pseudobulge, contributing to its growth \citep{Noguchi1999, Elmegreen2008, Mandelker2014}. However, this process can take up to $\sim 0.5$ Gyr, and if clumps are short-lived, they are disrupted faster than they reach bulge. It should be noted that bulge growth on cosmological timescales is well-established \citep{Hopkins2010, Brooks2016, Sachdeva2017}, although this process does not imply long-lived clump scenario by itself and can be possibly explained by other mechanisms, unrelated to clumps.

Summarising all previous findings, we can note that both clumps and spiral arms concentrate gas and star formation. At the same time, clumpy galaxies take place as the majority of star-forming ones instead of spiral galaxies at high redshifts, with occasional coexistence of two types of structure in a single galaxy. All this imply the possibility that clumpy galaxies may be the progenitors of spiral galaxies \citep[e.g.][]{Elmegreen2007, Margalef-Bentabol2022}. However, it is not well established if (and how) the interplay between spiral arms and clumps manifests itself more than simple simultaneous existence in a same galaxy. There are possible questions, for example, do clumps directly form into a spiral structure with time (for example, by the mechanism of swing amplification, \citealt{Julian1966, Goldreich1965b, Toomre1981}), or can spiral arms, in principle, stimulate clump formation.

Also, there are some clear differences between the properties of clumps and spiral arms, and they can influence the host galaxy in different ways. This means it is important to discern these objects, especially when they coexist, although it is rarely done. For example, in a recent work of \citet{Kalita2025a} it can be seen that a system of their detected clumps sometimes resemble spiral structure, so spiral arms' flux can possibly interfere with clumps' measurements. Other way, the studies of high-$z$ spiral structures could benefit from the separate consideration of clumps \citep[e.g][]{Chugunov2025a, Kalita2025c}. As a rare exception, \citet{Mercier2025} distinguish clumps and more extended substructures based on their surface area, but do not treat spirals specifically.

Considering all of this, in our work we aim to measure clumps' properties in high-$z$ galaxies, accounting separately for the existence of spiral structure. We are going to carefully discern the spiral pattern and clumps, avoiding the confusion of their fluxes. Fortunately, we have already done a decomposition of high-$z$ galaxies with the inclusion of spiral arms to model in \citet[][hereafter \citelinktext{Chugunov2025a}{Paper I}]{Chugunov2025a} and we can rely on its results in measuring clumps' properties. After doing this, we can compare the parameters of clumps and spirals directly, which can help us understand the connection between these two types of structure.

Throughout this paper, we assume a standard flat $\Lambda$CDM cosmology with $\Omega_m = 0.3, \Omega_\Lambda = 0.7$ and $H_0 = 70~\text{km}~\text{s}^{-1}~\text{Mpc}^{-1}$. All magnitudes are expressed in AB system \citep{Oke1974}.

\section{Data and methods}
\label{sec:data}

\subsection{Sample}
As mentioned before, in the current work we rely on our \citelinktext{Chugunov2025a}{Paper I} and the sample of galaxies to study is taken from it entirely. This sample consists of 159 galaxies with prominent spiral structure, located at $0.1 \leq z \leq 3.3$, imaged by HST and JWST. Specifically, there are 126 objects from the COSMOS survey done by HST \citep{Koekemoer2007}, where the selection of spiral galaxies was mostly done by \citet{Reshetnikov2022, Reshetnikov2023}, and these objects are located at $z \lesssim 1$. The remaining 33 galaxies at $z \gtrsim 1$ come from two JWST surveys: CEERS \citep{Bagley2023} and JADES \citep{Eisenstein2023, Rieke2023}. In these samples, galaxies were selected by hand and multi-wavelength imaging is available for them, allowing to study the dependence of any parameters on wavelength. For more information on the creation of this sample, we refer the reader to \citelinktext{Chugunov2025a}{Paper I}. We note that, unlike most studies of clumps, our sample includes only well-resolved spiral galaxies, which may bias our results.

\subsection{Clumps identification}
\label{sec:identification}
\begin{figure}
	\includegraphics[width=\linewidth]{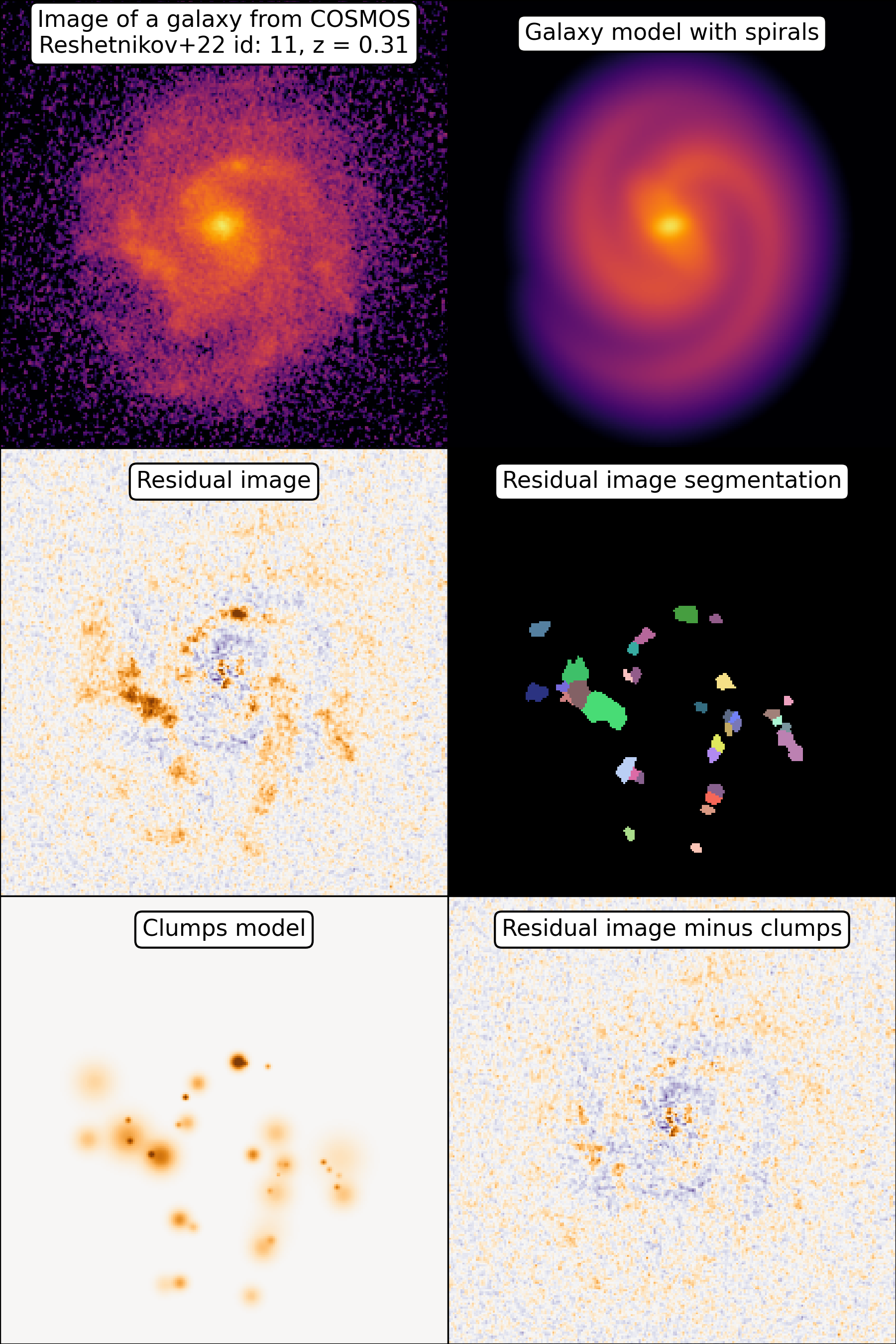}
	\caption{A schematic diagram showing the main steps in clumps' detection and parameters measurement. Top left: an example image of a galaxy from our sample. Top right: photometric model with spiral arms from  \citelinktext{Chugunov2025a}{Paper I}. Middle left: residual image from \citelinktext{Chugunov2025a}{Paper I} (image minus model), where orange represents positive values (model underestimates image), purple is negative and white is zero. Middle right: segmentation image of the residual, with each coloured blob showing a possible individual clump. Lower left: a photometric model of clump system, where each clump is modelled as a Gaussian convolved with PSF. Lower right: a residual image minus clumps model with the same colour map as the residual. Note that almost nothing except noise is remaining.}
	\label{fig:method_example}
\end{figure}

In \citelinktext{Chugunov2025a}{Paper I}, we have done a photometric decomposition including the spiral arms into a model, a method described in \citet{Chugunov2024}. As a result, we have a photometric model with spiral arms for each individual image, but these models do not account for clumps as the previous work was not focused on them. As a result, clumps appear as localised bright blobs on residual images (which is simply an image of a galaxy with its photometric model subtracted) and, given the otherwise good accordance between images and models, almost no other structures stand out from the background noise. This effect can be seen in Fig.~\ref{fig:method_example}.

Usually, clumps have to be detected in images of galaxies where clumps' flux is not yet separated from other components. The most common approaches rely on the observational fact that clumps are compact and thus have steep brightness gradients, so wavelet transform \citep{Sok2025} or subtracting a smoothed image \citep{Kalita2024} are widely used techniques to detect them. But the availability of residual images makes identifying clumps significantly easier and more straightforward than it has to be done in most works related to clumps. We can use one of methods designed to identify multiple objects on image.

As a first step, we apply \verb|photutils.segmentation| library for \verb|Python| \citep{Bradley2024}, specifically a function \verb|detect_sources| to residual images from \citelinktext{Chugunov2025a}{Paper I}. This function builds a segmentation image showing connected areas of higher brightness compared to background noise, i.e. it highlights clump candidates locations. For this step, we used a threshold of $1.5 \sigma + 0.001 F_m$ where $\sigma$ is a background noise and $F_m$ is a flux of galaxy model; the second term was added to discard false clump candidates arising from relative residuals at the centre of a galaxy. For robust segment detection, the residual image was additionally convolved with Gaussian kernel with a FWHM of 3 pixels. The minimal size of a segment was 20 adjacent pixels exceeding the threshold. This configuration was adopted as a result of trial-and-error process with visual control of segmentation images (see an example in Fig.~\ref{fig:method_example} where segment map covers all structures that visually appear as clumps on the residual image). Next, in order to separate adjacent clumps that may be labelled as a single large segment, we use \verb|deblend_sources| function from the same library. It is designed to perform deblending based on multi-thresholding and watershed segmentation. We apply an additional morphological deblending step. As we assume that individual clumps are not highly elongated (especially after PSF smearing), we perform a distance transform for each segment. We treat peaks on distance map as the locations of individual clumps, splitting each segment when needed between these peaks. The final segmentation map (with an example shown in Fig.~\ref{fig:method_example}) represents the locations and number of individual clumps, and is used in next steps for precise estimation of their parameters.

\subsection{Clumps measurement}

For the next step, we use a photometric decomposition package \verb|IMFIT| \citep{Erwin2015}, same as in \citelinktext{Chugunov2025a}{Paper I}. Our goal here is to fit each individual clump on residual images with a symmetric 2D Gaussian function, thus the best-fit parameters describe its position, size and luminosity. Segmentation maps obtained in a previous step were used to produce initial guess for fitting. Residual images to fit are taken from \citelinktext{Chugunov2025a}{Paper I}, and the same noise maps and PSF are used. We use the Levenberg --- Marquardt algorithm \citep{More1978, Markwardt2009} which is used in \verb|IMFIT| by default.

We choose to not use a more versatile and general S{\'e}rsic function \citep{Sersic1968} for fitting, because many clumps lie on the edge of resolution, making measurement of their S{\'e}rsic indices questionable. We also do not allow Gaussian to be asymmetric (i.e. having elliptical isophotes) as clumps in residuals seem to be mostly circular and this would also introduce additional parameters. Therefore, each individual clump adds 4 parameters to the model. For galaxies which have multi-wavelength imaging, we set the number and positions of clumps fixed between filters, to avoid any systematics related to the different image quality. Specifically, we fit clumps' parameters including positions in F200W of F210M filter (whichever is available), count clumps with fitted non-zero brightness and, for other bands, fit them with their positions fixed.

An additional detail of the fitting process is that we choose to do it iteratively. As clumps can be numerous, it could be time-consuming to fit all them at once, because Levenberg --- Marquardt algorithm becomes significantly slower with the number of free parameters increasing. However, clumps are mostly separated spatially, so fitting their parameters simultaneously brings little benefit. Therefore, at each iteration we add a single clump (or a small group of clumps if they are located close to each other) to a model and fit their parameters; for next iteration, these parameters are set fixed and next clump or small group is added, and so on.

As a result, we obtain a photometric model of clumps system, with an example shown in Fig.~\ref{fig:method_example} and similar images for all galaxies are available at \url{https://github.com/IVChugunov/Distant_spirals_decomposition}. For each clump, we have information of its position, size and luminosity directly from their fitted parameters. For clumps in galaxies from the JWST subsample, the variation of parameters with wavelength is also available.

\begin{figure*}
	\includegraphics[width=\textwidth]{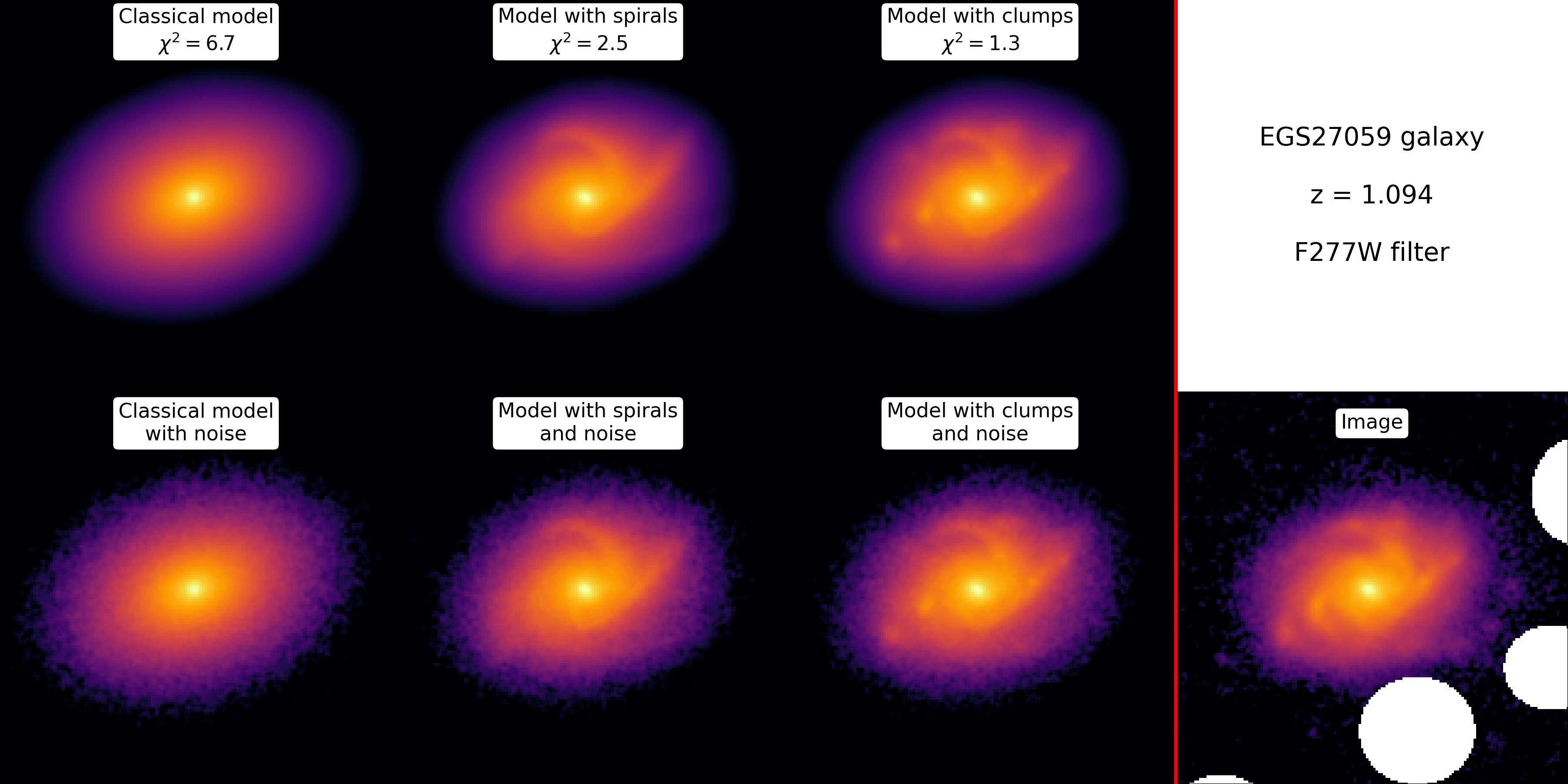}
	\caption{A collage showing an example galaxy image (rightmost) and its models (three remaining columns). Top row shows original models. Bottom row shows the same models with superimposed noise of the same magnitude as in original image for visual comparison with it. From left to right: a classical model (bulge + disc) from \citelinktext{Chugunov2025a}{Paper I}, a model including spiral arms from \citelinktext{Chugunov2025a}{Paper I} and a model with spiral arms with added clumps model from this work. $\chi^2$ statistic is also shown for each model.}
	\label{fig:models_overview}
\end{figure*}

As one can see in Fig.~\ref{fig:method_example}, residual images after clumps subtraction retain almost no bright or large-scale structures. One can say that model with spiral arms from \citelinktext{Chugunov2025a}{Paper I} summed with clumps model provides an almost complete photometric description of a galaxy. As an example of this, in Fig.~\ref{fig:models_overview} we show the set of models for a single image, including the model with spirals and clumps. The model with clumps with added noise is remarkably similar to the image of the real galaxy.

It should be noted that somewhat similar approach was used in \citet{Kalita2025a}, although they fitted clumps over simple bulge + disc model, which does not allow one to discern clumps and spiral arms, if the latter are present. Authors of mentioned work were used asymmetric Gaussians to fit clumps, but in some cases they tend to fit spiral segments instead of individual clumps --- see their Figs. 4 and 5.

Of course, our method is not universal and it does not invalidate other approaches. It is useful in the specific case studied where we are fortunate enough to have models with spiral arms already. Otherwise, preparing them from scratch would be a time-consuming process (although spiral arms models bring a lot of information themselves). Note that spiral arms may be absent in the majority of galaxies in other works.

\subsection{SED fitting}
\label{sec:sed_fitting}
Spectral energy distribution (SED) fitting is a widely used and powerful technique to measure properties of stellar population and, possibly, associated interstellar medium, based on their multi-wavelength photometry. The essence of this method is to fit a theoretical spectrum of stellar population to the set of flux observations in multiple different filters. In turn, theoretical spectrum of a stellar population with interstellar medium depends on their properties, such as stellar mass, age, star formation rate, metallicity, dust content and so on. Therefore, these parameters can be potentially retrieved with SED fitting \citep[see a review by][]{Conroy2013}.

As for some galaxies we have clumps' properties measured in different filters, it is natural to use this data for SED fitting. There are some works where this procedure is applied, such as \citet{Guo2015, Sok2025, Kalita2025a}. For this purpose, we use \verb|Prospector| library for \verb|Python| \citep{Johnson2021} which works with \verb|FSPS| stellar population synthesis library \citep{Conroy2010}, includes various optimisation methods and convenient functions for actual SED fitting. Different studies of clumps apply various SED fitting packages, and \verb|Prospector| was used in a very recent study in \citet{Zhu2026}.

\begin{figure*}
	\includegraphics[width=\textwidth]{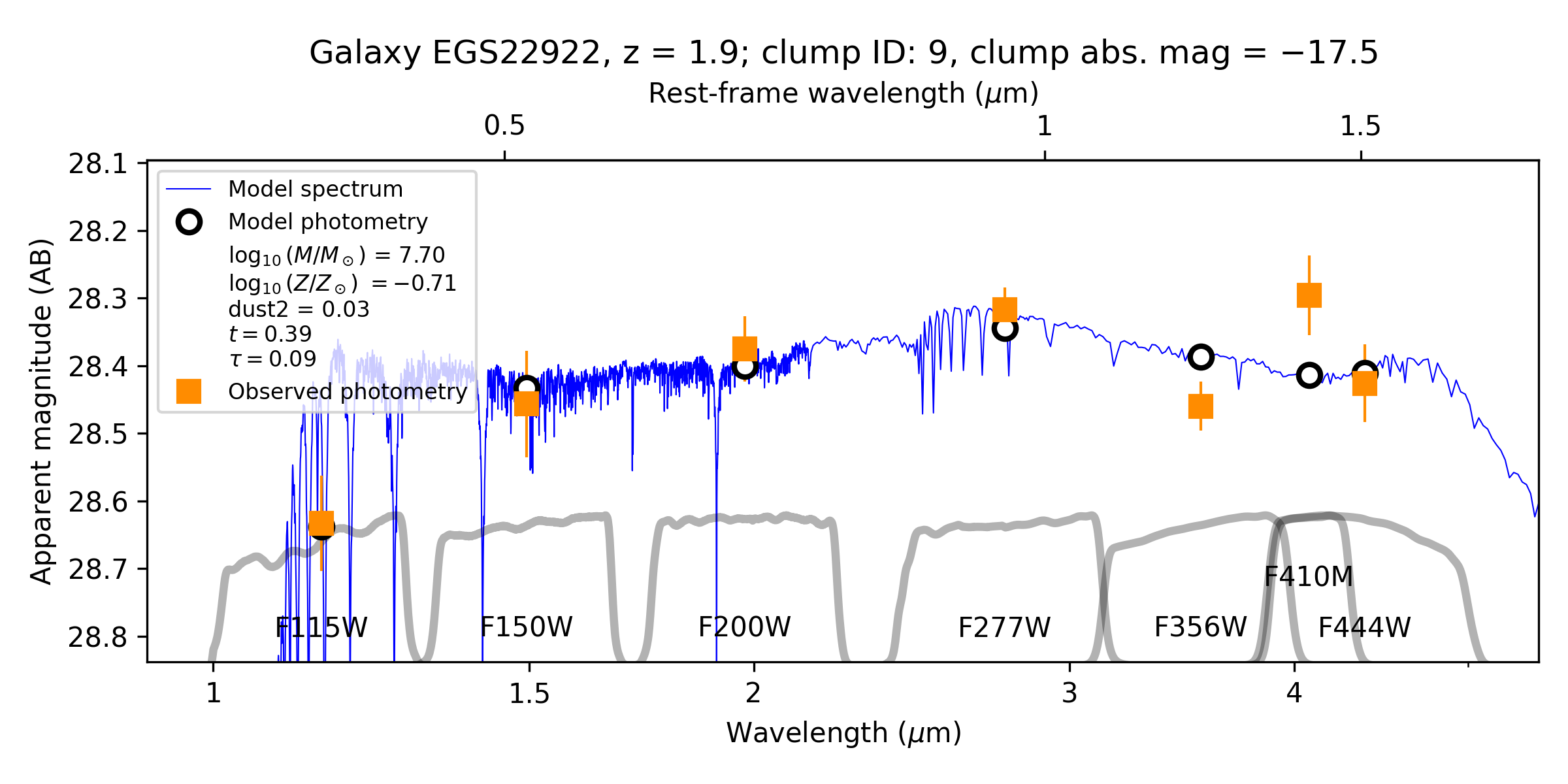}
	\caption{An example of SED fitting result using Prospector. Observed fluxes are shown in squares, each one corresponds to a single filter; their passbands are shown in bottom. The model spectrum is shown as a thin curve, and corresponding model photometry is represented by circles. The parameters of the fit \citep[see][]{Conroy2010, Johnson2021} are shown in the legend.}
	\label{fig:sed_fit}
\end{figure*}

Therefore, we select bright clumps having measurements in a wide range of wavelengths and perform SED fitting. As our wavelength coverage is limited and we risk encountering degeneracies between parameters, we employ Markov Chain Monte Carlo (MCMC) analysis to estimate the reliability of our measurements. We use parametric star formation history (SFH) in a form of delayed exponential ($\text{SFR} \sim t \times \exp(-t/\tau)$) as we aim to recover only the most general properties of SFH. We note that clumps tend to have lower metallicity than the host galaxy. In general, their metallicities are believed to be solar or sub-solar \citep{Estrada-Carpenter2025}, and dust attenuation is moderate \citep{Elmegreen2009, Bassett2017, Kalita2025b}, not exceeding 2 mag in $V$ band in most cases. Accordingly, we limit possible $\log [Z/Z_{\sun}]$ to $-1.0\ldots0.0$ range, and $A_V$ to $0.0\ldots2.0$ mag. In Fig.~\ref{fig:sed_fit}, we show an example of SED fitting for a single clump.

\section{Results}
\label{sec:results}

\begin{figure}
	\includegraphics[width=\linewidth]{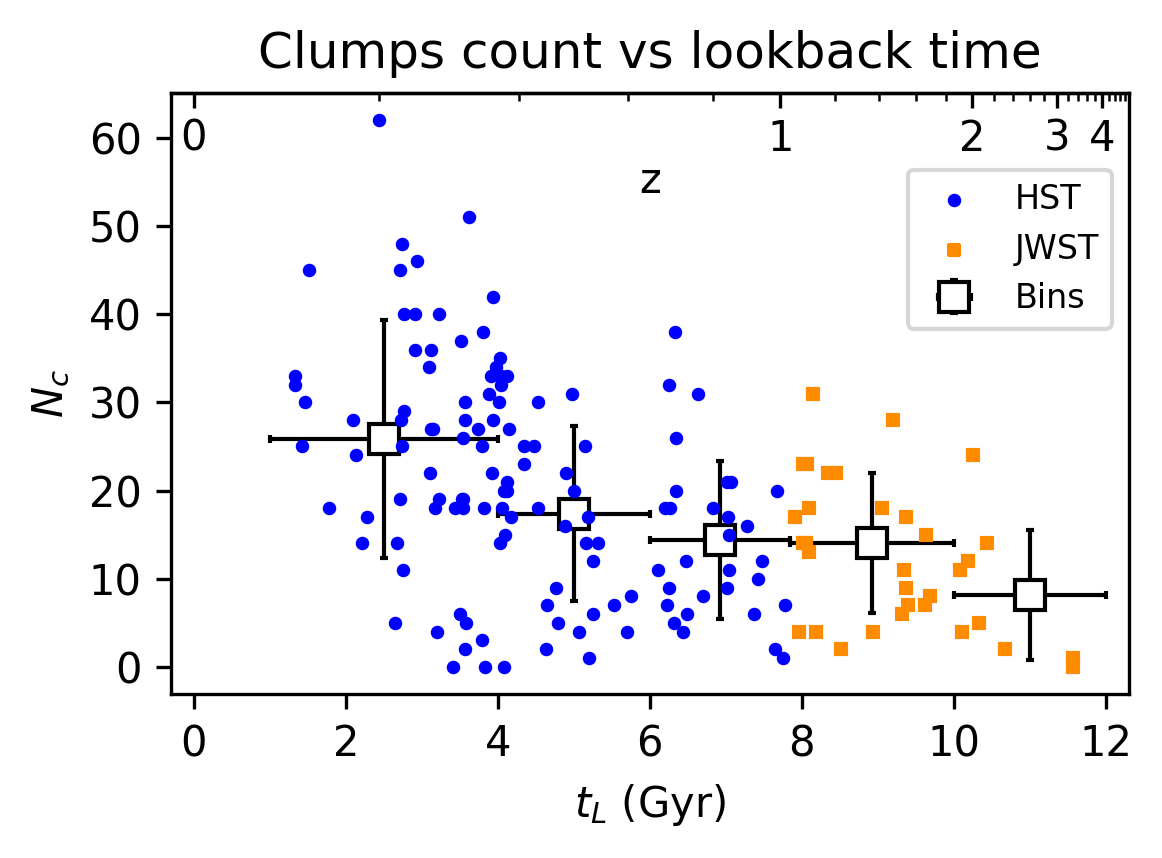}
	\caption{The dependence of total clumps count in a galaxy $N_c$ on lookback time $t_L$. Black squares with error bars represent binned averages, with horizontal bars representing bin ranges and vertical standing for standard deviations.}
	\label{fig:N_time}
\end{figure}

After we applied our method of clumps identification and measured their parameters, we found a total of 3003 clumps, or 19 per galaxy on average. Only 4 galaxies out of 159 have no clumps detected with implied threshold. We see less clumps in galaxies with increasing distance, as shown in Fig.~\ref{fig:N_time}. As we will show in Section~\ref{sec:CT}, it is heavily influenced by decreasing image quality with $z$. For 33 galaxies from JWST subsample, the combined number of clumps is 411 (12 per galaxy); they represent $z > 1$ range, and multi-wavelength data is available for them. To compare with, \citet{Kalita2025a} detected 167 clumps in 32 galaxies at $z \sim 1.5$; \citet{Guo2018} found 3193 clumps in 1270 galaxies at $0.5 \leq z < 3$ (although this was done in pre-JWST era). The fact that we usually see more clumps per galaxy may be attributed to the properties of our sample (as described in \citelinktext{Chugunov2025a}{Paper I}, we initially selected well-resolved galaxies to study spirals, likely biased towards higher masses). Another possible reason is methodological differences: photometric decomposition with spiral models likely reproduces the smooth structures of galaxy more accurate than commonly used automated methods, allowing us to detect fainter objects. Finally, two-stage deblending procedure (see Section~\ref{sec:identification}) may additionally increase the number of detected clumps (see Fig.~\ref{fig:method_example}).

\subsection{Clumps' luminosities and observational effects}
\label{sec:CT}
Clumps contribution to the total luminosity of a galaxy (hereafter clump-to-total ratio, or $C/T$, where ``total'' is counted as a sum of model components' luminosities) is one of parameters of our interest. We can found it directly from the decomposition model including clumps and other components. For example, its dependence on $z$ is another way to examine how the ``clumpiness'' of galaxies varies with time. We now mention that $C/T$ observed ``as is'' have mean value of nearly 4.5\% and highest of 16\%, but before doing any meaningful analysis of $C/T$, one has to account for possible observational effects.

We can make a correction for observational effects in a similar way that we done in \citelinktext{Chugunov2025a}{Paper I}. The most obvious observational effect is the varying image quality, already mentioned before. First, it is known that the spatial resolution decreases with increasing distance to the object, at least up to $z \approx 1.7$ \citep[for details, see][]{Melia2018, Whitney2020}. Also, the cosmological dimming leads to the decrease of surface brightness proportional to $(1 + z)^4$ \citep[for the details and derivation, see][]{Calvi2014, Whitney2020}, and both these effects influence our ability to detect clumps. Indeed, \citet{Mercier2025} report that high-redshift substructures often remain undetected. They claim that, despite apparent clumpy fraction in near-infrared was increasing with time from $z = 4$ to $z = 1$, the intrinsic trend was actually the opposite.

Secondly, we have to account for possible band-shifting effects. If galaxies at different redshifts are observed in the same filter, the rest-frame wavelengths of actually received radiation vary. It can distort any results if the observed quantity varies with redshift itself. One can expect band-shifting effects to influence clumps much, because these objects are indeed brighter in shorter wavelengths \citep[e.g.][]{Calabro2019}, in accordance with their active star formation.

Finally, our sample is not complete in mass: as spiral galaxies are not as abundant at high redshift, we did not cut our sample down even more, but this makes massive galaxies represented at high-$z$ domain more than in low-$z$. Regarding clumps, \citet{delaVega2025} reported that the fraction of clumpy galaxies correlates with total stellar mass, although \citet{Guo2015, Sattari2023} found the anti-correlation between the said parameters. Finally, \citet{Huertas-Company2020} claims that, the more massive galaxies appear clumpy more often, but clumps' contribution to the stellar mass is lower. Therefore, it is important to account for galaxies' masses while analysing $C/T$.

\begin{figure}
	\includegraphics[width=\linewidth]{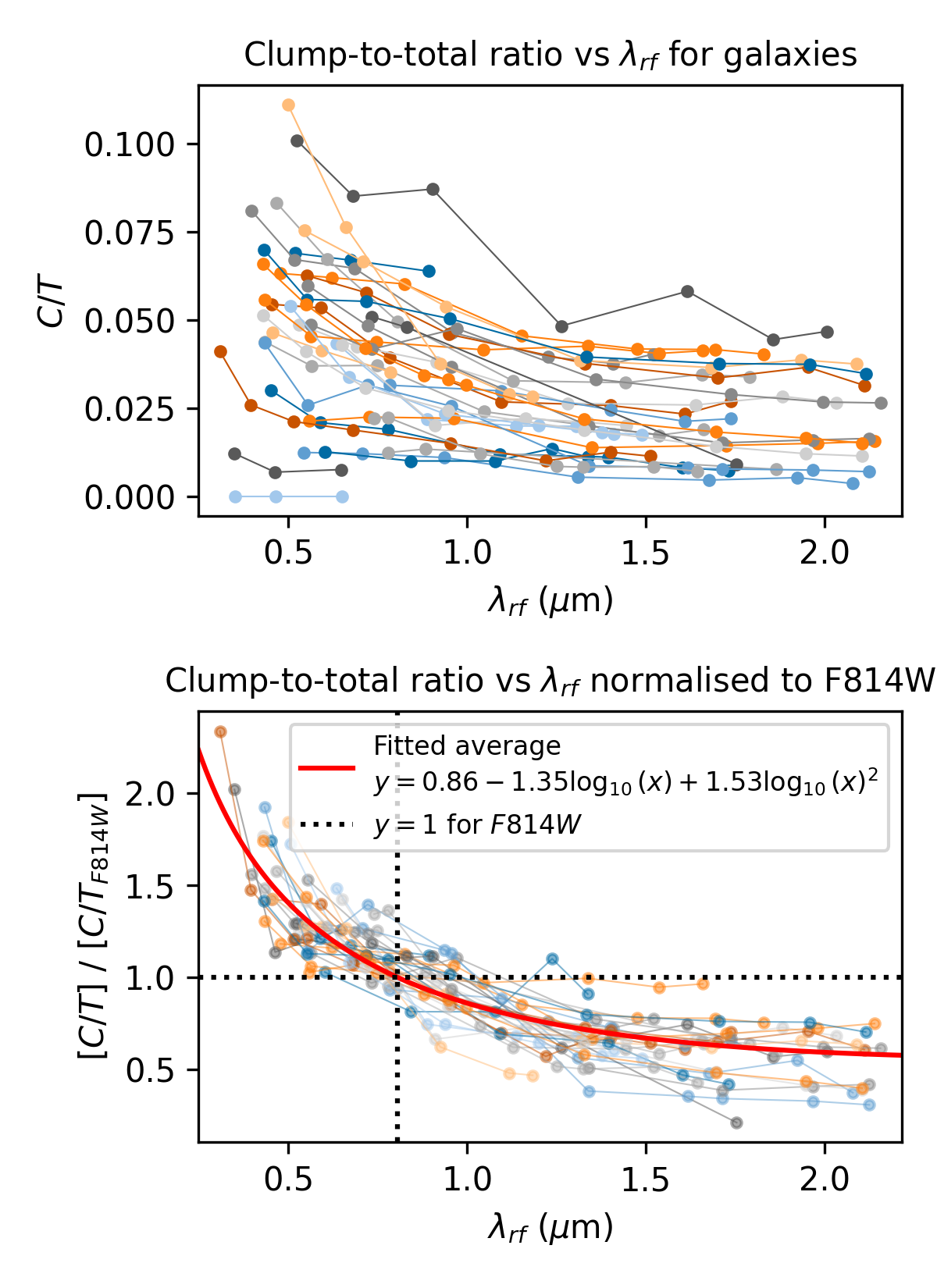}
	\caption{Top: the dependence of clump-to-total ratio $C/T$ on rest-frame wavelength $\lambda_\text{rf}$ for individual galaxies from JWST subsample. Bottom: $C/T$ normalised to the same level (corresponding to a unity for rest-frame wavelength of $F814W$ filter) for different galaxies. Normalization is done by fitting the same square logarithmic function (shown as a thick line) with normalisation coefficients individual for each galaxy}
	\label{fig:CT_wavelength}
\end{figure}

The most straightforward effect to account for is band-shifting, as we have a subsample of objects measured in multiple filters. In Fig.~\ref{fig:CT_wavelength}, we show how $C/T$ changes with rest-frame wavelength $\lambda_\text{rf}$ for them. One can observe that $C/T$ increases significantly towards shorter rest-frame wavelengths, in a similar way for different galaxies, following the trend in \citet{Messa2022}\footnote{Note, however, that in the work of \citet{Guo2012} no difference was found between rest-frame $U$ and $V$ contributions.}. Then we select a single anchor wavelength for a galaxy as nearly 806 nm. This is a pivot wavelength $\lambda_p$ of F814W \citep[see][]{Hathi2024} which is a filter used for COSMOS subsample, and for the majority of galaxies in JWST subsample it is inside the range of wavelengths covered by a set of filters. Next, we are going to ``reduce'' $C/T$ of all galaxies to this value, introducing $C/T_\text{F814W}$. For the majority of JWST galaxies, it can be done by interpolating their $C/T$ to the said wavelength. However, some galaxies lack the required wavelength coverage: in particular, there is only a single filter available for the COSMOS subsample, so we calculate $C/T_\text{F814W}$ for them assuming their $C/T(\lambda_\text{rf})$ follows the same law as objects from JWST subsample. We fit our data from Fig.~\ref{fig:CT_wavelength} with a square logarithmic function in the form of $y = a \log_{10}(\lambda)^2 + b \log_{10}(\lambda) + c$ with a normalisation constants corresponding to each galaxy individually, thus approximating $C/T(\lambda_\text{rf})$ for them. This approximation is shown in the bottom of Fig.~\ref{fig:CT_wavelength} as $[C/T] / [C/T_\text{F814W}]$ and we use it to recalculate $C/T$ observed in $\lambda_\text{rf}$ to $F814W$ wavelength. Note that, compared to $F814W$ filter, $C/T$ is almost twice smaller in near-infrared (1.5\ldots2 $\mu$m), and nearly twice as large at the edge of ultraviolet (0.4 $\mu$m). After doing this correction, we see that the highest and the average values for $C/T_\text{F814W}$ slightly drop to 11\% and 3.6\%, respectively.

\begin{figure}
	\includegraphics[width=\linewidth]{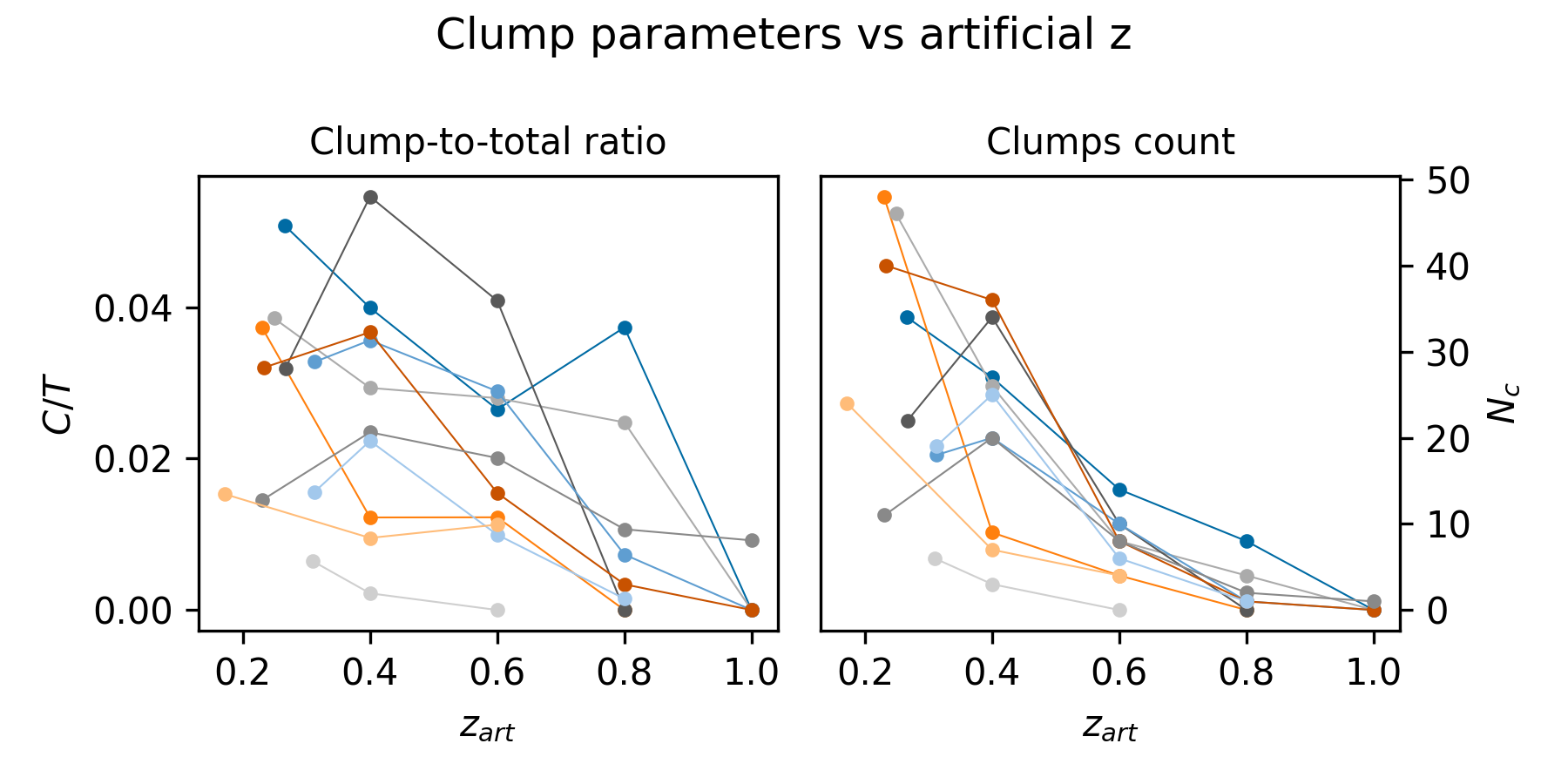}
	\caption{The dependence of measured clump parameters for individual galaxies artificially redshifted to different $z_\text{art}$. Left: clump-to-total ratio $C/T$. Right: total number of clumps in a galaxy $N_c$. The leftmost point for each line is true $z$ of a galaxy.}
	\label{fig:CT_art_z}
\end{figure}

Now, we are going to explore how observational effects impact our measurements. We cannot observe the same galaxy at different redshifts, but instead we can do artificial redshifting \citep[which is a commonly used technique: see, e.g.][]{Martinez-Garcia2023, Kuhn2024}. It is done by selecting nearby galaxies and decrease the quality of their images as if they were observed at some higher artificial redshift $z_\text{art}$. After that, we apply the same methods to measure clumps in artificially redshifted images and if measured parameters turn out to be different at different redshifts, it means that image quality influences the results. We have used this approach in \citelinktext{Chugunov2025a}{Paper I} (see Section 5.2 there and Fig. 17 in particular) and now we use the same artificially redshifted images as prepared in \citelinktext{Chugunov2025a}{Paper I}. In Fig.~\ref{fig:CT_art_z} we show how $C/T$ and $N_c$ changes for individual galaxies with artificial redshifting. It is seen that decreasing image quality mostly decreases observed $C/T$ as well as the number of detected clumps. In distant galaxies, faint features are more likely to remain undetected, and tight groups of clumps can appear as a single object. However, this conclusion is qualitative, and we will account for varying image quality in a different way following in this Section.

\begin{figure*}
	\includegraphics[width=\textwidth]{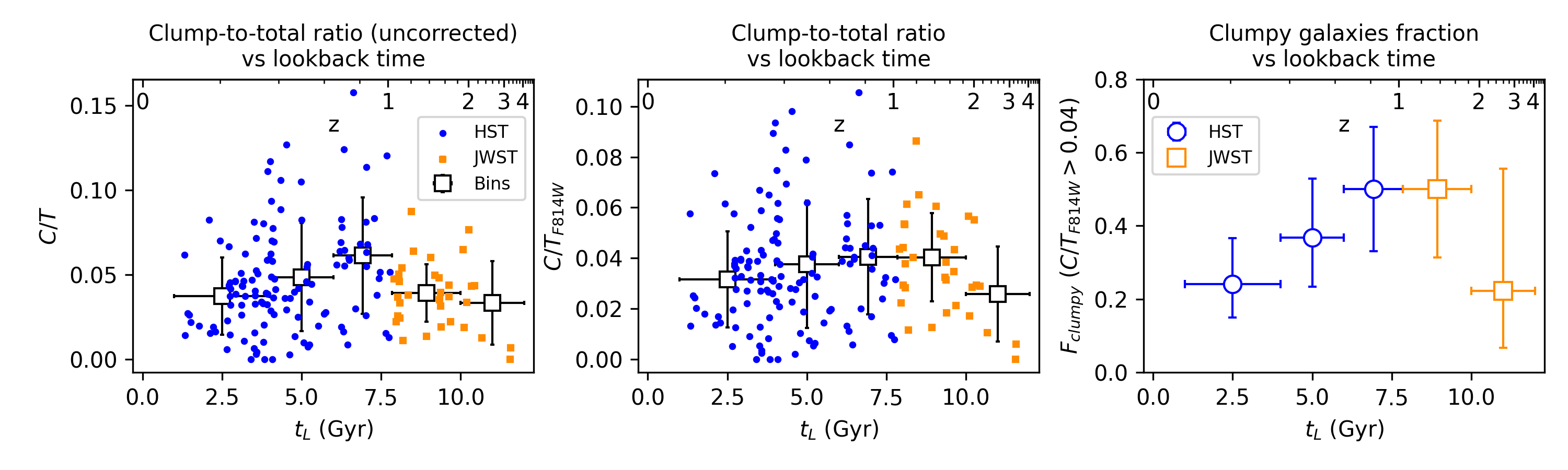}
	\caption{Left: clump-to-total ratio $C/T$ versus lookback time $t_L$ for individual galaxies. Middle: the same, for clump-to-total ratio corrected to represent the F814W filter, $C/T_\text{F814W}$. Right: the fraction of galaxies with $C/T_\text{F814W} > 0.04$ in the same bins. Here, vertical error bars represent the 95\% confidence intervals.}
	\label{fig:CT_true_z}
\end{figure*}

We cannot explore the dependence of $C/T$ on mass or luminosity in the same simple way, and we will do this following in this Section as well. Now we focus on the dependence of $C/T_\text{F814W}$ on observed redshift of a galaxy $z$. In Fig.~\ref{fig:CT_true_z}, we show $C/T$ and $C/T_\text{F814W}$ for individual galaxies as well as values averaged in lookback time bins. We also show the fractions of clumpy galaxies in the same bins $f_\text{clumpy}$, assuming that $C/T_\text{F814W} > 0.04$ is a condition to count galaxy as clumpy. First, we compare diagrams for $C/T$ (which is a value not corrected for rest-frame wavelength) and a corrected value $C/T_\text{F814W}$. One can see that $C/T$ values behave discontinuously at $z \approx 1$ where HST part of our sample ends and JWST part begins. This discontinuity represents that HST and JWST filters have different wavelengths, leading to the change of observed $C/T$; the same effect appeared in \citelinktext{Chugunov2025a}{Paper I} (see, e.g. Fig.~12). When wavelength correction is applied, we get a $C/T_\text{F814W}$ parameter which does not show any discontinuity at $z \approx 1$, validating our approach. Further on, we disregard $C/T$ and consider only a corrected value $C/T_\text{F814W}$.

We see that peak in both $C/T_\text{F814W}$ and $f_\text{clumpy}$ is located at $z \sim 1\ldots2$. Parameter $f_\text{clumpy}$ is commonly measured in literature, and the trend observed in our work is broadly consistent with it \citep[e.g.][]{Murata2014, Shibuya2016, Sattari2023, delaVega2025, Sok2025}, although the exact values of $f_\text{clumpy}$ vary from study to study and seem to depend on host galaxy mass and selected threshold. Nevertheless, its peak at $z \sim 1\ldots2$ roughly coincides with the peak of star formation in the Universe, also known as ``cosmic noon'' \citep{Madau2014}. Similarly, the peak of molecular gas density occurs roughly at the same time \citep{Walter2020}. As clumps are star-forming structures, the connection between these parameters can be expected.

\begin{figure*}
	\includegraphics[width=\textwidth]{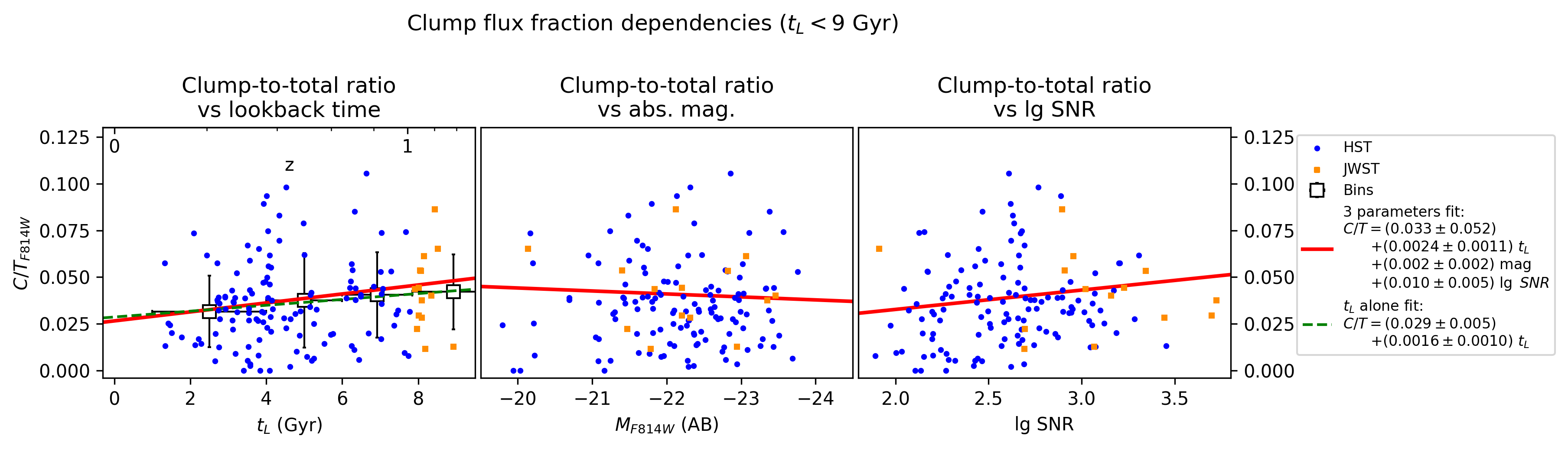}
	\caption{The dependence of $C/T_\text{F814W}$ on three parameters. From left to right: lookback time $t_L$ (limited to $t_L < 9$ Gyr), absolute magnitude of the host galaxy $M_\text{F814W}$ and signal-to-noise ratio of a galaxy $\lg \text{SNR}$. Red line shows the projections of a trilinear function fitted jointly to $C/T_\text{F814W}$ depending on all three parameters, to each single parameter. For comparison, green dashed line shows a single linear fit to $C/T_\text{F814W}$ as a function of $t_L$ only.}
	\label{fig:CT_three_pars}
\end{figure*}

Finally, regarding the dependence of image quality and the incompleteness of our sample, we are going to explore the dependence of $C/T_\text{F814W}$ on these parameters jointly with $z$ (or lookback time $t_L$). We will do this in the same manner as we done in \citelinktext{Chugunov2025a}{Paper I} to disentangle the dependencies of different spirals' parameters on various properties of a galaxy (see Section 4.3 in the mentioned work). We select the luminosity of a galaxy $M_\text{F814W}$ as a somewhat reasonable proxy for mass to account for the incompleteness of our sample. As an image quality measure, we take the signal-to-noise ratio inside the disc of a galaxy (SNR); both these parameters were already introduced in \citelinktext{Chugunov2025a}{Paper I}. Finally, we fit $C/T_\text{F814W}$ as a trilinear function of $t_L$, $M_\text{F814W}$ and $\lg \text{SNR}$, as shown in Fig.~\ref{fig:CT_three_pars}. Note that in Fig.~\ref{fig:CT_true_z} we observed $C/T_\text{F814W}$ increases up to $z \sim 1\ldots2$, so we limit this analysis only to $t_L < 9$ Gyr. We conclude that $C/T_\text{F814W}$ depends on SNR as expected (better SNR allows one to detect more clumps) and slightly increases with lookback time up to $z \sim 1\ldots2$, with a rate of 0.2\% per Gyr. We observe no $C/T_\text{F814W}$ variation with absolute magnitude beyond errors. The results in literature regarding this matter are inconclusive \citep{Guo2015, Huertas-Company2020, Sattari2023, delaVega2025}. As brighter and more massive galaxies are likely to have better SNR, it is possible that mass dependency is also influenced by image quality dependency. In any case, our findings apply only to spiral galaxies which also may influence results.

\begin{figure}
	\includegraphics[width=\linewidth]{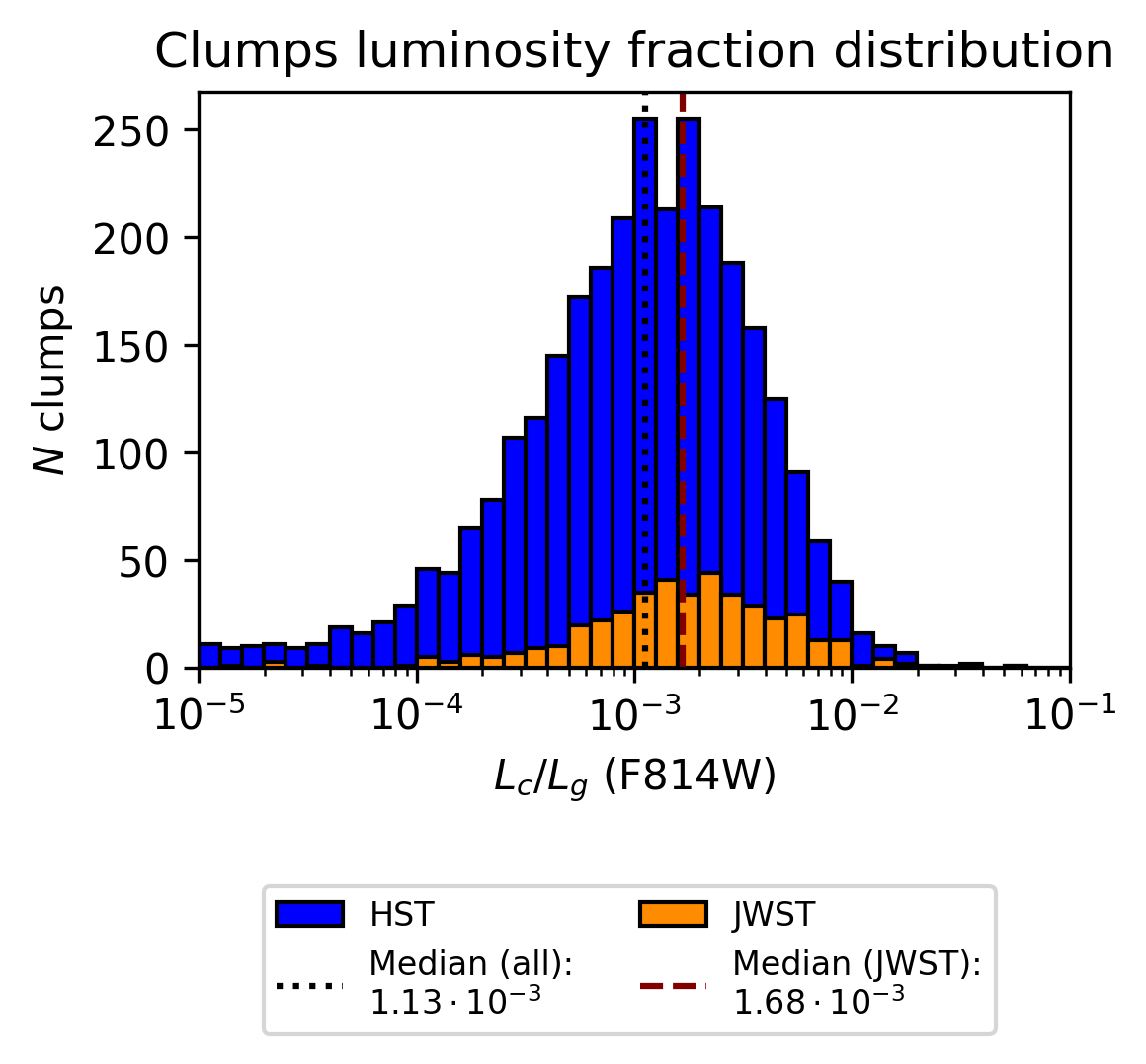}
	\caption{A stacked histogram of clump contribution to the total luminosity of a galaxy $L_c/L_g$ normalised to $\lambda_p$ of F814W filter. The median value for JWST subsample is shown.}
	\label{fig:LcLg_hist}
\end{figure}

Regarding individual clumps, we consider the ratio between clump luminosity $L_c$ and galaxy luminosity	$L_g$. We show the distribution of $L_c / L_g$ normalised to the $\lambda_p$ of F814W filter in Fig.~\ref{fig:LcLg_hist}. The highest $L_c / L_g$ in the entire sample is 6\%. The median $L_c / L_g$ in our entire sample is $1.1 \times 10^{-3}$, and for JWST subsample $(z > 1)$ it is $1.7 \times 10^{-3}$.

\begin{figure}
	\includegraphics[width=\linewidth]{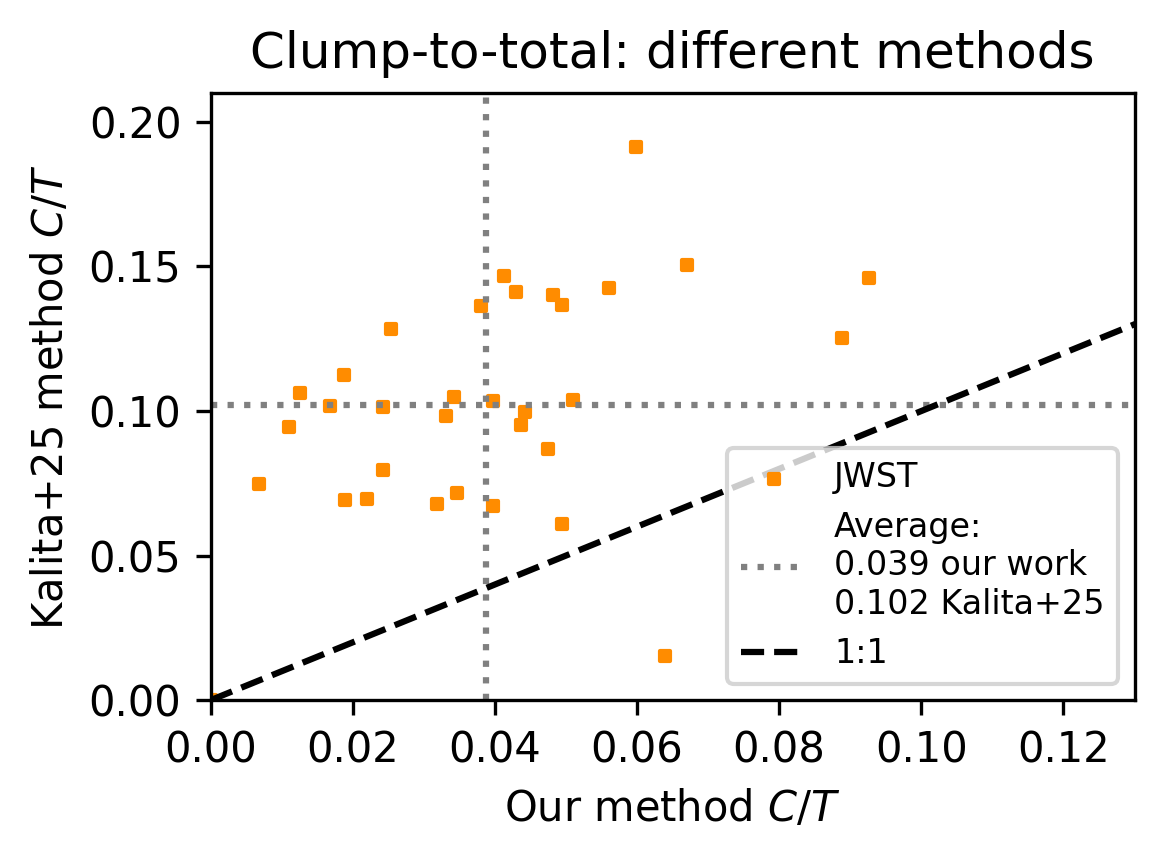}
	\caption{A diagram showing the comparison between $C/T$ of galaxies from JWST samples, measured with different methods: the one from our work, and another from \citet{Kalita2025a}. 1:1 line is shown, as well as average values.}
	\label{fig:methods_comp}
\end{figure}

Combining the results in this Section together, we can note that our $C/T_\text{F814W}$ and $C/T$ are significantly smaller than values found in the literature. For example, \citet{Kalita2025a} found that combined contribution of clumps to the total mass of galaxy can reach 20\%, and for SFR it can reach 30\%. Meanwhile, the highest $C/T_\text{F814W}$ in our data is 11\%; although this is not a direct comparison, but the radiation of this part of the spectrum consists of both young and old stellar population, thus being related to both mass and SFR. Moreover, we see that $C/T$ at $\lambda_\text{rf} \approx 2 \mu$m does not exceed 5\% for all galaxies which have sufficient filter coverage. At this wavelength, it is commonly assumed that different stellar populations have almost the same mass-to-light ratio \citep{McGaugh2014}, and therefore it is safe to assume that clumps contribution to the total mass of galaxy does not exceed 5\%. One possible explanation is related to methods: in our work, spiral arms are modelled separately, while neglecting to do this could lead to the attribution of spiral arms flux to clumps. As highest $S/T$ in our sample approach 60\%, we can estimate that, if spiral arms are not treated properly, then roughly one-third of their flux can be wrongly ascribed to clumps. To find how significant this effect is, we consider the method from \citet{Kalita2025a} to locate and find clumps. In short, in their method clumps are fitted to residual image from bulge+disc model, and detected in image where smoothened image was subtracted. We reproduce and apply it for our galaxies and compare the obtained $C/T$ with the results of our method. In Fig.~\ref{fig:methods_comp}, we show this comparison for our JWST subsample.

It can be seen that method from \citet{Kalita2025a} almost always yields higher $C/T$, demonstrating that methodological differences are crucial. If $C/T$ values are measured with their method, the average (10\%) and highest (nearly 20\%) values become approximately consistent with \citet{Kalita2025a}, even though the sample is different. It is also important that our sample consists entirely of galaxies with established spiral structure but it is not the case of most other studies. In turn, the formation of spiral structure requires disc to be partially stabilised \citep{Dekel2009}, which may be unfavourable for massive clumps formation. This may even mean that we are exploring different structures compared to those commonly called ``clumps''. However, as samples of galaxies in other works include spiral galaxies as well, they are not entirely different from ours. Notably, \citet{Mercier2025} found that substructures' (including clumps) contribution to the total flux of a galaxy near rest-frame 1 $\mu$m does not exceed a few per cent, but this value was measured using a very high threshold for clump detection, needed to ensure detection uniformity up to $z = 4$.

The same difference is found for individual clump contributions to the luminosity of a galaxy $L_c / L_g$. \citet{Kalita2025a} reports that individual clumps can contribute up to 10\% to their host galaxy mass and 30\% to their SFR; in \citet{Sok2025} data, some clumps have luminosity exceeding 10\% in all redshift bins. In our data, the highest $L_c / L_g$ in any filter is only 6\%. Regarding the median values, \citet{Sok2025} report that, the majority of clumps have $L_c / L_g$ around $10^{-2}$ at $1.2 \leq z < 2$, and we find the median value of $1.7 \times 10^{-3}$ in the comparable redshift range (although they consider rest-frame UV luminosity, where clumps are brighter). Most likely, the same explanation as for overall $C/T$ is responsible for the lack of brighter clumps in our sample. Either spiral galaxies do not host clumps as bright as clumpy ones do, or $L_c / L_g$ may be overestimated due to spirals' flux attribution to the clumps. Concerning the lower $L_c / L_g$ values overall, they are possibly also linked to other methodological reasons. First, the photometric decomposition model of a galaxy with spiral arms may approximate the smooth part of its light distribution more accurately than different automated methods, and the resulting residuals may be cleaner, allowing us to detect fainter structures over the noise. Also, as explained by \citet{Dessauges-Zavadsky2017}, clump clustering may artificially increase the luminosities of clumps, as multiple clumps may be counted as one. However, we apply deblending of clumps while identifying them (see Section~\ref{sec:identification}), so this issue may be mitigated in our work, at least partially. Therefore, our method may have higher sensitivity, but as a tradeoff, it is accompanied with the difficulty to implement.

\subsection{Clumps' luminosities and other parameters of galaxies}
\label{sec:CT_misc}
\begin{figure}
	\includegraphics[width=\linewidth]{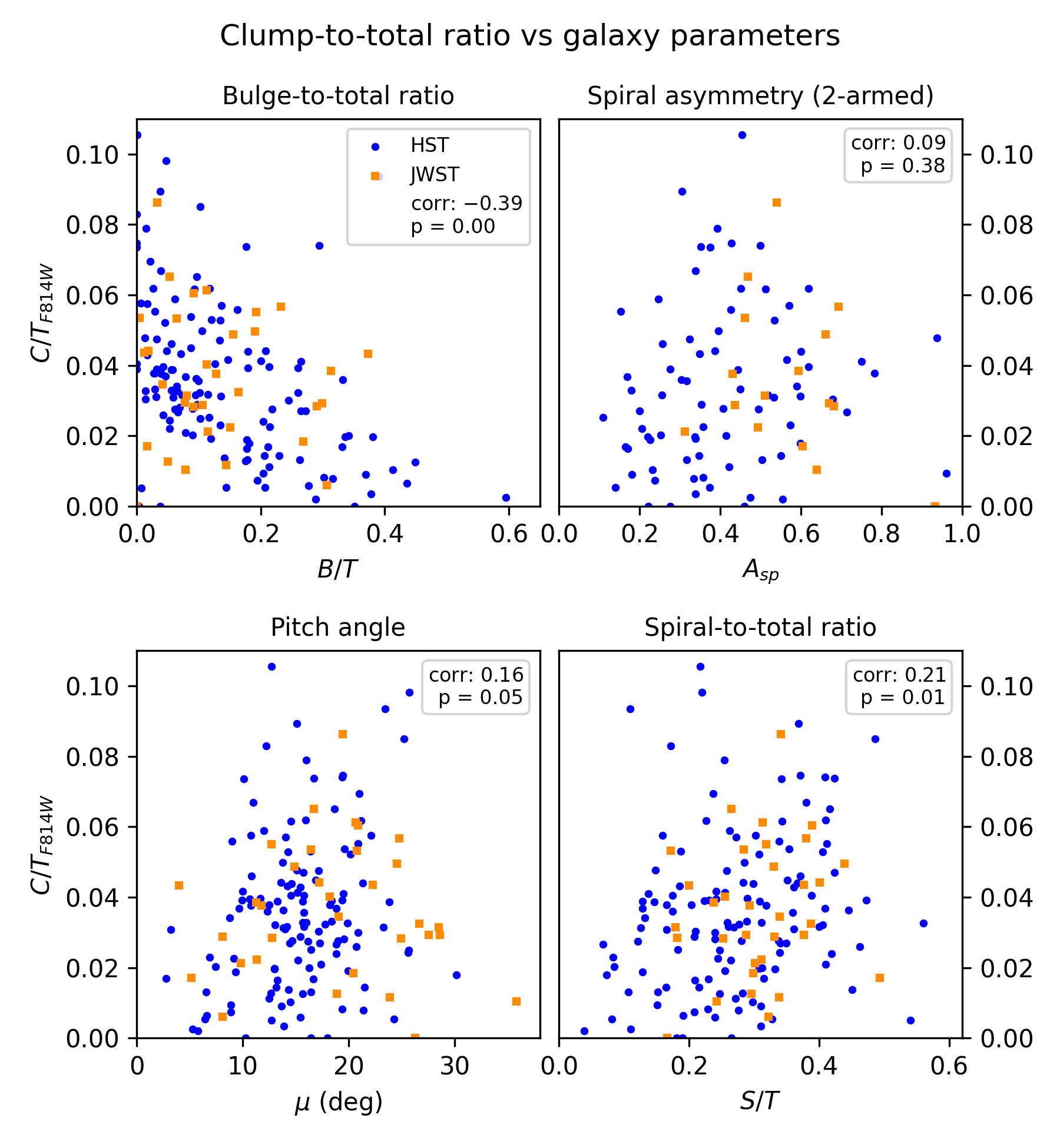}
	\caption{Diagrams of $C/T_\text{F814W}$ depending on different parameters of a galaxy. Top left: bulge-to-total ratio $B/T$. Top right: asymmetry index of the spiral structure $A_\text{sp}$ for two-armed spirals. Bottom left: average pitch angle $\mu$ of a galaxy. Bottom right: spiral-to-total ratio $S/T$. In the legend, Pearson correlation coefficient and p-value are shown.}
	\label{fig:CT_params}
\end{figure}

Now we focus on some other parameters. We examine how $C/T_\text{F814W}$ is connected with different parameters of the host galaxy. First, we consider a diagram of $C/T_\text{F814W}$ versus bulge-to-total ratio $B/T$, shown in Fig.~\ref{fig:CT_params}. We see a statistically significant anti-correlation between these parameters: galaxies with bright bulges almost never have prominent clump system. It agrees with previous observations \citep{Kalita2024} and aligns with the theoretical consideration that massive bulges stabilize discs and prevent the formation of clumps (at least, long-living ones, see \citealt{Dekel2023}). Note that the connection between bulge and spiral arm properties is also widely discussed in literature \citep[e.g.][]{Kennicutt1981, Seigar2005a, Yu2019}.

Next, we turn to the possible connection between $C/T_\text{F814W}$ and parameters of the spiral structure: the spiral-to-total ratio $S/T$, spiral structure asymmetry $A_\text{sp}$ and average pitch angle $\mu$. All these parameters were measured in \citelinktext{Chugunov2025a}{Paper I}. The corresponding diagrams are shown in Fig.~\ref{fig:CT_params}. We observe weak positive correlation between $C/T_\text{F814W}$ and $S/T$. The most prominent feature of this diagram is the lack of galaxies with low $S/T$ and high $C/T_\text{F814W}$. We believe this is a selection effect: if a galaxy has weak spirals and multiple luminous clumps, its spiral structure becomes difficult to be recognized, and it is not included in our sample of spiral galaxies. There is no statistically significant correlation between $C/T_\text{F814W}$ and $A_\text{sp}$: the most clumpy galaxies are not ones with the most asymmetric spiral structure, and vice versa. In \citelinktext{Chugunov2025a}{Paper I} we found that spiral structure in high-$z$ galaxies tends to be more asymmetric compared to low-$z$ ones, and we have drawn a hypothesis that it is connected to the higher clumpiness of high-$z$ galaxies. However, the lack of strong correlation between $C/T_\text{F814W}$ and $A_\text{sp}$ is an evidence against that idea, although we do not challenge higher clumpiness of high-$z$ galaxies. It seems that high spiral asymmetry of high-$z$ galaxies should have another origin. For example, galaxies in the early Universe had a higher frequency of tidal interactions \citep{Lavery2004, Conselice2022} which could lead to an increased perturbation of spiral structures within discs; a good example of such perturbation in the local Universe is M~51 \citep{Marchuk2024b}. Finally, the correlation between $C/T_\text{F814W}$ and $\mu$ is weakly positive, at the edge of significance. Probably, both these parameters can be influenced by $B/T$: we observe weak anti-correlation between $\mu$ and $B/T$ in our data with Pearson coefficient $-0.19$ (although both were measured in \citealt{Chugunov2025a}, we did not compare them directly). Such a correlation aligns with the density wave theory \citep{Lin1964} which predicts that higher mass concentration should produce more tightly wound spirals in a galaxy, and $B/T$ is a reasonable proxy for stellar mass concentration. Observational studies are divided over this matter: in some works, $\mu$ is reported to negatively correlate with $B/T$ \citep[e.g.][]{Yu2019, Chugunov2024} or related parameters, such as bulge mass or luminosity, concentration parameter or shear rate \citep[e.g.][]{Seigar2005a, Savchenko2013}, and some do not find it \citep[e.g.][]{Kendall2015, Hart2017}. Possibly, various proxies may approximate mass concentration with different accuracy, and also different samples of galaxies may have not the same dominant spiral formation mechanisms.

\subsection{Clump sizes}
\label{sec:sizes}
\begin{figure}
	\includegraphics[width=\linewidth]{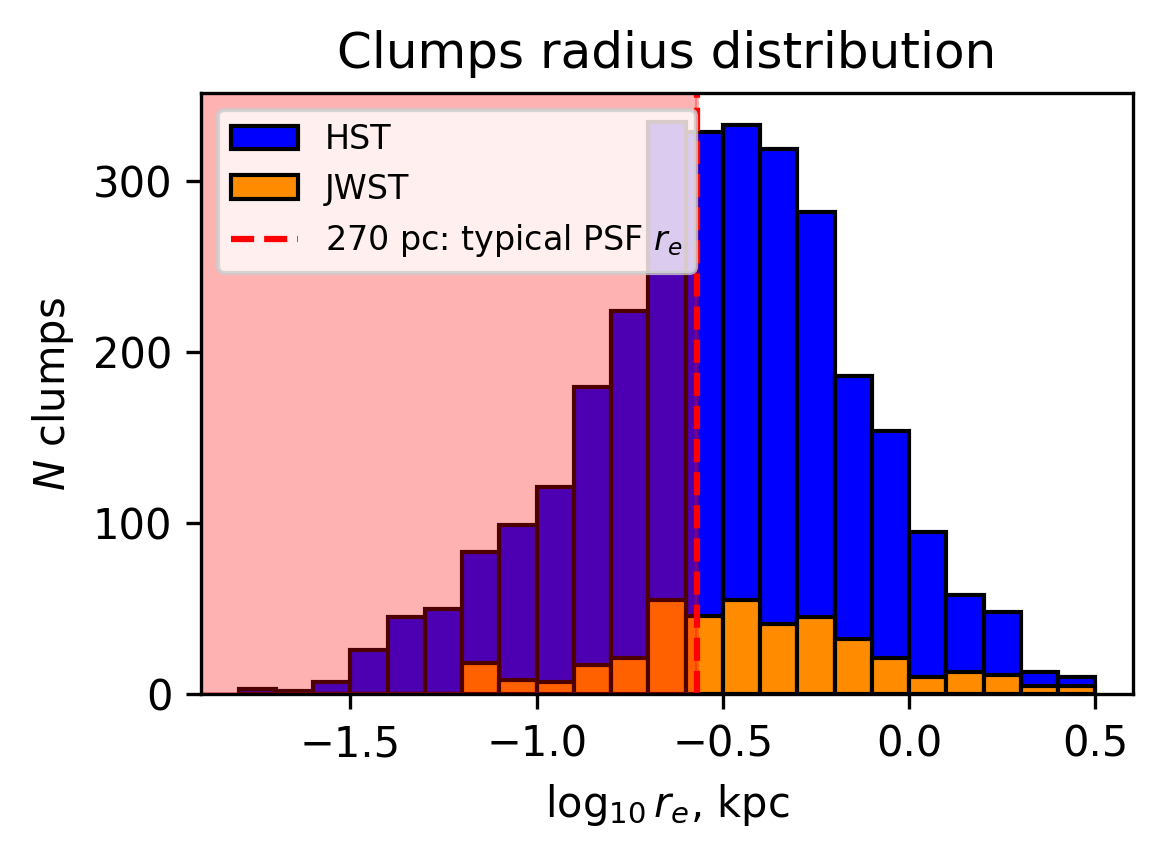}
	\caption{A stacked histogram of logarithm of effective radii $r_e$ for individual clumps. Red shaded area on the left represents $r_e < 270$ pc, which is the typical size of PSF, so their exact measurements are not highly reliable.}
	\label{fig:r_hist}
\end{figure}

In this section, we examine the measured sizes of clumps in our sample. We use effective radius $r_e$ (which contains a half of the total clump luminosity) as a measure of clump size, and in Fig.~\ref{fig:r_hist} we show a distribution of individual clumps by $r_e$. Note that many clumps have their intrinsic sizes smaller than PSF. For instance, PSF $r_e$ for F814W filter corresponds to 270 pc in linear size at $z = 0.45$ which is average $z$ for galaxies in COSMOS subsample. If measured size of some clump is smaller, we cannot treat it as reliable and only consider it unresolved. We observe that our sizes of clumps are slightly smaller than in other works. In \citet{Kalita2025a}, the median size of clump is about 500 pc (see Fig.~10 in their work) and their galaxies are located at $z = 1.5$. In our data, taking only the JWST subsample (which galaxies are located at $z > 1$ and have average $z = 1.5$) yields the median clump size of 360 pc. The reasons for this difference can possibly be similar to those for clump luminosities (see discussion in Section~\ref{sec:CT}). Firstly, they can be possibly attributed to different method of measurement: in our work, we have subtracted spiral structure previously, and the light from spirals cannot be attributed to clumps. Secondly, the difference between spiral and clumpy galaxies may manifest itself. \citet{Elmegreen2009} found that clump sizes relative to the host galaxy bulge size are smaller in spiral galaxies than in clumpy ones (although they notice that spiral galaxies are characterised by prominent bulges, and clumpy ones are not).

\begin{figure}
	\includegraphics[width=\linewidth]{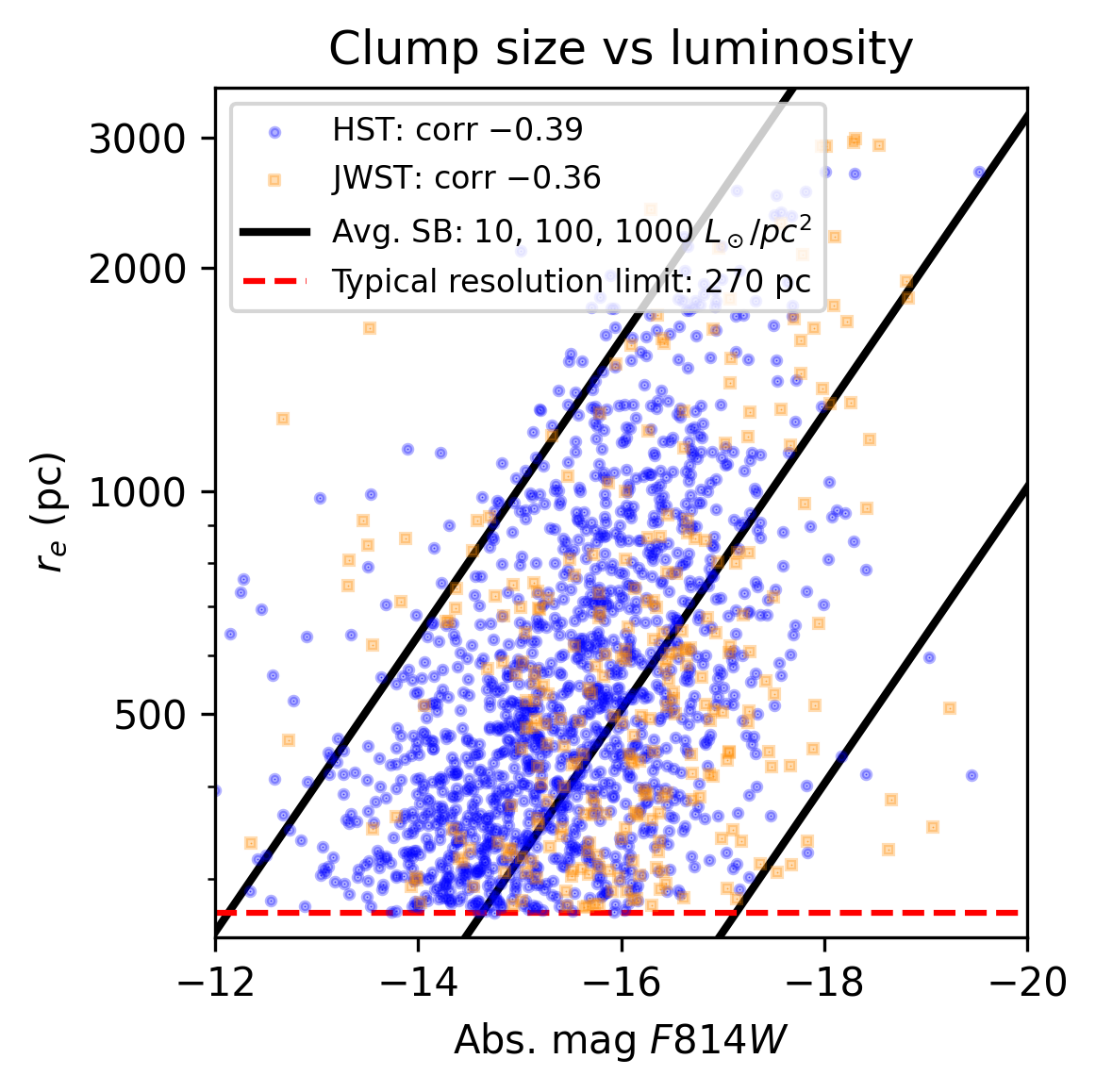}
	\caption{Clumps $r_e$ versus their absolute magnitude $M_\text{F814W}$. Lines of constant average surface brightness are also shown: from left to right, they correspond to 10, 100 and 1000 $L_{\sun} / \text{pc}^2$. Effectively unresolved clumps ($r_e < 270$ pc) are not shown.}
	\label{fig:r_mag}
\end{figure}

Next, we compare the sizes of clumps and their luminosities. We show a diagram of these two values in Fig.~\ref{fig:r_mag}. We observe a correlation between their absolute magnitude and effective radius, which is expected. For example, the dependence between clumps' mass and size was observed in different works \citep[see, e.g.][]{Cava2018, Messa2022, Claeyssens2023, Kalita2025a}, reproduced in simulations \citep{Mandelker2017} and similar relation exists for young star clusters in the local Universe \citep{Brown2021}. The vast majority of clumps have their average surface brightness (inside $r_e$) being between 10 and 300 $L_{\sun} / \text{pc}^2$.

\subsection{Spatial distribution of clumps}
\label{sec:location}
\begin{figure}
	\includegraphics[width=\linewidth]{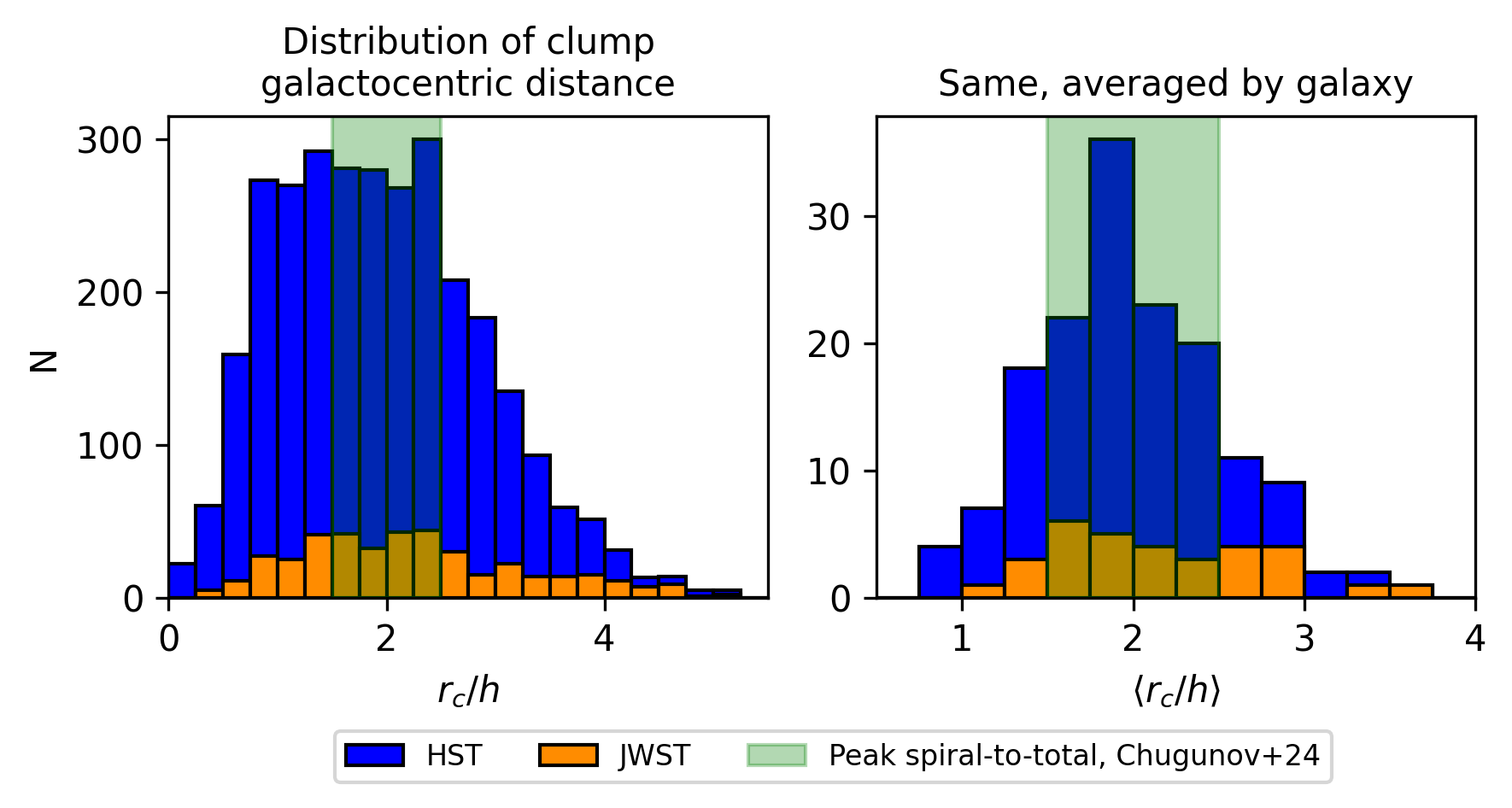}
	\caption{The distributions of galactocentric distance $r_c$ of individual clumps, normalised to disc exponential scale $h$. Left: a stacked histogram of $r_c / h$ for individual clumps. Right: a stacked histogram of $\langle r_c / h \rangle$ for individual galaxies in our sample, i.e. $r_c / h$ averaged over entire galaxy (weighted average by clump luminosity was used). Green shaded area represents the range of galactocentric radii where spiral contribution to the luminosity of a galaxy reaches its peak, according to \citet{Chugunov2024}, specifically 1.5--2.5 $h$.}
	\label{fig:rch}
\end{figure}

In this Section, we focus on spatial distribution of clumps in their host galaxy. First, we are interested in their radial distribution based on galactocentric radius $r_c$. In order to compare the results for different galaxies of different sizes, we normalise $r_c$ to the exponential scale of galactic disc $h$ and examine $r_c / h$, accounting for disc orientation parameters measured in \citelinktext{Chugunov2025a}{Paper I}. In Fig.~\ref{fig:rch}, we show the distribution of this value for individual clumps as well as averaged values for the entire galaxies $\langle r_c / h \rangle$. We see that clumps tend to appear mostly at galactocentric radii of 1--2.5 $h$. Remarkably, in \citet{Chugunov2024}, we found that spiral arms contribution to the galaxy brightness profile is highest at 1.5--2.5 $h$, so clumps and spiral arms tend to appear in the same regions of the disc. This is expected, as star formation defines both clumps and spiral arms. In turn, star formation occurs in disc and requires gas density to be sufficiently high, which possibly limits the range of galactocentric radii of both spiral arms and clumps. In any case, now we have shown this based on observations.

Now we examine further if locations of clumps coincide with spiral arms. \citet{Kalita2025a} mentions that 70\% of clumps overlap with residual spiral features; their analysis was based on visual inspection, and we are going to explore this in a more robust way. We first outline the fundamental idea: consider a galaxy and assume that spirals occupy some fraction of its disc $x$ (for example, in image pixels). Then, if clumps are distributed independently from spiral pattern, the fraction of clumps located in spirals out of total number of clumps, $f_\text{cs}(x)$ should be nearly equal to $x$. Otherwise, it would point to some sort of connection between clumps and spirals.

\begin{figure}
	\includegraphics[width=\linewidth]{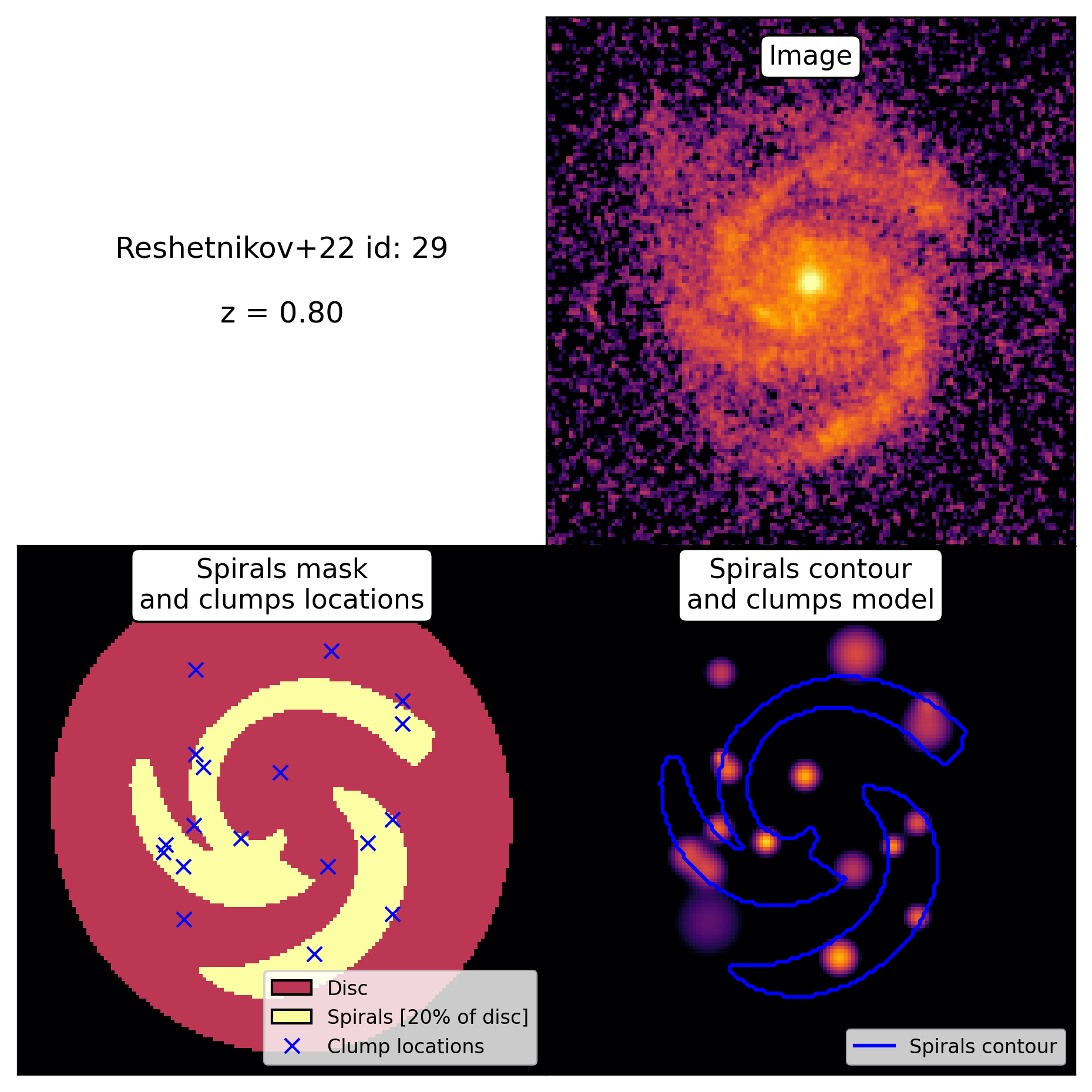}
	\caption{A scheme showing the process of separation spiral arms and inter-arm area, and exploring the spatial distribution of clump centres. In this case, spirals are assumed to occupy 20\% of the disc. Top right: the image of an example galaxy. Bottom left: separation of spiral arms and the remaining disc, and the locations of clumps. Bottom right: the photometric model of clump system and superimposed contour of spiral structure.}
	\label{fig:clump_loc_scheme}
\end{figure}

To implement this idea, one has to define how to select which pixels belong to spiral arms and which ones are not, i.e. construct a spiral mask. Spiral arms detection on image is a non-trivial task \citep[see, e.g.][]{Davis2014, Forgan2018, Bekki2021, Querejeta2021, Walmsley2023}, but we already have models with spiral arms from \citelinktext{Chugunov2025a}{Paper I}, and we can use them. We assume that pixels with the highest brightness of spiral models are related to spiral structure, so we have to select some brightness threshold. To compare different galaxies uniformly, we select a threshold in a way that spiral mask occupies some fixed fraction of disc in all galaxies: for example, 20\%. The disc itself is defined as the area inside $4h$ from the centre of a galaxy. In Fig.~\ref{fig:clump_loc_scheme}, we show an example of such separation.

\begin{figure}
	\includegraphics[width=\linewidth]{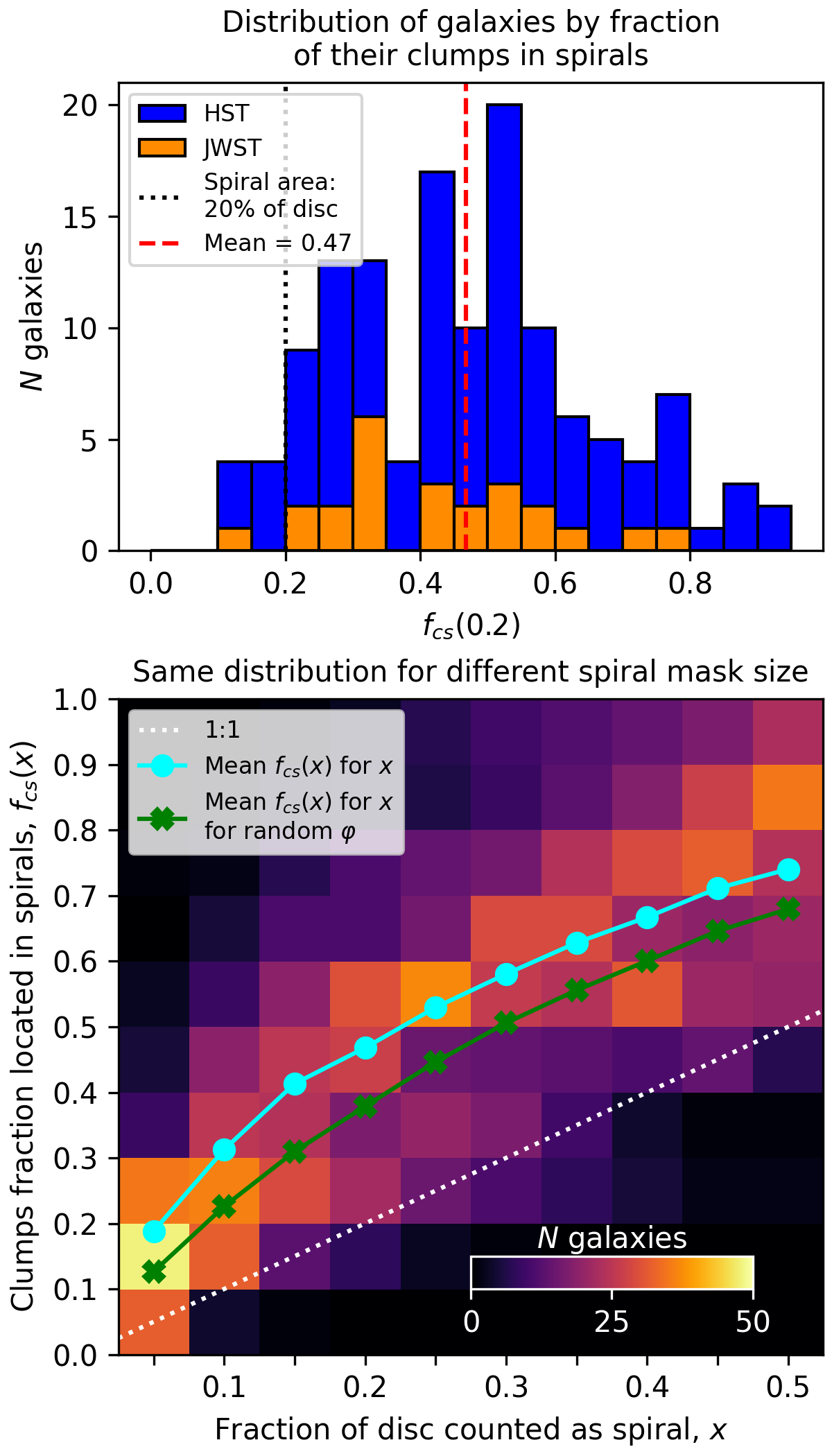}
	\caption{Top: a stacked histogram of the fraction of clumps located in spiral arms $f_\text{cs}(0.2)$ for individual galaxies with at least 5 detected clumps. Here, spiral mask is defined as 20\% disc pixels with highest brightness of spirals. Bottom: the heat map showing the distributions of $f_\text{cs}(x)$ for different $x$ (each column represents a histogram for single $x$). Average $f_\text{cs}(x)$ values are also shown. Cyan circles show true $\langle f_\text{cs}(x) \rangle$, green crosses represent $\langle f_\text{cs}(x) \rangle$ if clumps were distributed randomly by azimuth, but retained their radial distribution (see text).}
	\label{fig:clump_loc_hist}
\end{figure}

After this is done, we count a number of clumps which centres are located in spiral mask, and calculate a fraction of such clumps out of total. For meaningful statistics, we consider only 133 galaxies containing at least 5 detected clumps. Thus, we obtain $f_\text{cs}(0.2)$ for them. Further on, we will describe this parameter as clumps concentration towards spirals. The distribution of this value is shown on the top of Fig.~\ref{fig:clump_loc_hist}, and in most cases $f_\text{cs}(0.2) > 0.2$, with an average value is 0.47. Of course, our decision to label exactly $x = 0.2$ of the disc area as spiral mask is arbitrary. In order to validate our result, we perform the same analysis, adopting multiple different $x$ values, as shown on the bottom of Fig.~\ref{fig:clump_loc_hist}. We observe that, regardless of $x$, it holds that $f_\text{cs}(x) > x$. This means that the fraction of clumps located in spiral mask is higher than the fraction of disc area occupied by spiral mask. In other words, clumps are concentrated towards spiral arms. Note that \citet{Foyle2010} made a very similar plot for different tracers of mass, stars and SFR, see their Fig. 2.

We test this result by slightly altering our method, and qualitatively it holds the same. First, we can define spiral mask in a different way: we select pixels with with highest spiral-to-total ratio instead of highest brightness of spiral arms. Next, we can define $f_\text{cs}(x)$ calculating the fraction of clumps flux in spiral mask instead of number of clumps. In any case, $f_\text{cs}(x) > x$ regardless of $x$, which shows that clumps are concentrated towards spiral arms. However, for $x = 0.2$ we have average $f_\text{cs}(0.2) = 0.46$ which is smaller than 70\% reported by \citet{Kalita2025a}; the similar $f_\text{cs}(x)$ is achieved only at $x = 0.45$. Although the choice of $x$ is arbitrary, the resulting spiral masks correspond to a visual impression of spirals location at $x = 0.2$ well enough (see Fig.~\ref{fig:clump_loc_scheme}), and look too extended with $x = 0.45$. Therefore, we observe lesser (although still significant) concentration of clumps towards spiral arms than \citet{Kalita2025a}. We may explain this difference by the fact that they did not account for spiral arms while preparing images to clump identification. They were detecting clumps based on residual images from bulge + disc model, but for spiral galaxies this approach leads to an over-subtraction in inter-arm region and under-subtraction in spiral arms \citep[see, e.g.][]{Chugunov2024, Marchuk2024b, Marchuk2025}. Consequently, the residual image has negative brightness in interarm region, and fainter clumps located there may remain undetected, leading to the under-representation of them.

Then, the next question arises. We already found that the radial distributions of clumps and spirals are similar (see Fig.~\ref{fig:rch}); couldn't it be the sole reason of this apparent concentration? When a question formulated this way, it is reasonable to examine the azimuthal distribution of clumps. To do this, we employ the following idea. Each clump has its own polar galactocentric coordinates $(r_c, \varphi_c)$; then we change azimuthal coordinates for each clump, assigning them randomly, so coordinates change to $(r_c, \varphi_\text{rand})$. We call this set an ``azimuthally random distribution''. If this operation does not change $f_\text{cs}$ at all, it means that the apparent concentration of clumps towards spiral arms is not connected to their azimuthal distribution. Otherwise, azimuthal distribution also matters. In Fig.~\ref{fig:clump_loc_hist}, we show $f_\text{cs}$ as a function of $x$ for azimuthally random distribution, comparing it with the values for real clump distribution. We see that the randomisation of $\varphi_c$ decreases the concentration of clumps towards spirals, but some degree of concentration remains: for example, $f_\text{cs}^{\text{rand}}(0.2) = 0.38$, varying by less than 0.01 for different random realisations. It means that clumps not only populate the same radii in the galactic disc as spirals, but also concentrate to spiral arms themselves.

\begin{figure}
	\includegraphics[width=\linewidth]{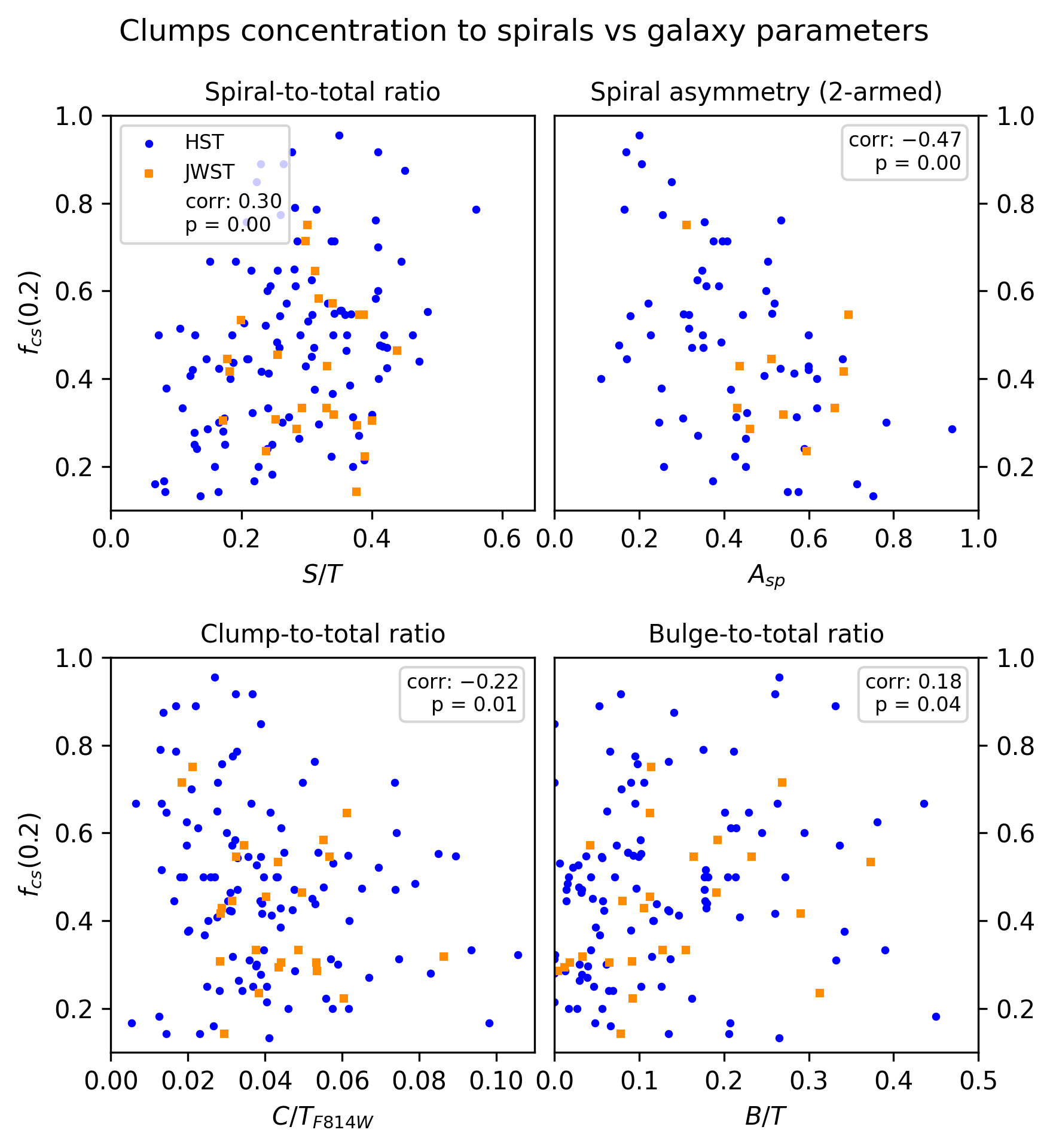}
	\caption{Diagrams of clumps concentration towards spirals $f_\text{cs}(0.2)$ dependence on different parameters of a galaxy. Top left: spiral-to-total ratio $S/T$. Top right: asymmetry index of the spiral structure $A_\text{sp}$ for two-armed spirals. Bottom left: clump-to-total ratio $C/T_\text{F814W}$. Bottom right: bulge-to-total ratio $B/T$. In the legend, Pearson correlation coefficient and p-value are shown.}
	\label{fig:fcs_params}
\end{figure}

Next, we examine how clumps concentration towards spirals $f_\text{cs}(0.2)$ depends on other parameters of the host galaxy. In Fig.~\ref{fig:fcs_params}, we show the diagrams for a few parameters where formally significant (albeit not very strong) correlation is observed. For $f_\text{cs}(0.2)$, we observe a positive correlation with spiral-to-total ratio $S/T$ and, to a lesser extent, bulge-to-total ratio $B/T$. There is a negative correlation with clump-to-total ratio $C/T_\text{F814W}$ and the asymmetry index of spiral structure $A_\text{sp}$. The correlation with $S/T$ is the most natural to expect, as brighter spirals are, by definition, generally concentrate more light and mass. The relations with $C/T_\text{F814W}$ and/or $A_\text{sp}$ may indicate that the most clumpy and/or asymmetric galaxies have clumps not collected in spiral arms, but scattered across the disc. We believe this represents the fundamentals of VDI \citep{Dekel2009}: unstable discs can form clumps everywhere, they are more clumpy and their spiral structure, if exist, is likely less ordered and symmetric. Finally, the weak positive correlation with $B/T$ is likely a manifestation of that massive bulge contributes to the stabilization of the disc and suppressing clumps formation (see also our Fig.~\ref{fig:CT_params}).

\subsection{Spectral energy distribution in clumps}

The radiation of a galaxy at different wavelengths is produced by its different populations, mainly young and old stars as well as various fractions of interstellar gas and dust. Therefore, studying the spectral energy distribution (SED) of a galaxy or its individual components helps us to understand their physical properties. For 33 galaxies in our JWST subsample we have images in multiple filters covering wavelength range from blue to near-infrared (see, e.g. Fig.~\ref{fig:CT_wavelength}). In this Section, we will focus on this information.

\subsubsection{Colour indices}
Colour index is a simple and commonly used indicator of spectral energy distribution which is normally counted as a difference of object magnitudes in two filters. However, our sample contains galaxies located at different $z$, and colour index derived from two fixed filters will represent SED slope at different wavelength, which is not very useful. Instead, we can approximate a rest-frame $g - r$ colour of the object, interpolating the magnitudes in a set of filters to a wavelengths of $g$ and $r$ bands ($\lambda_p = 477$ and 622 nm respectively, \citealt{Fukugita1996}). This can be done not only for entire galaxies, but for separate model components as well. Of course, this approach is not highly precise and it does not account for filter bandpass and SED details, and we should keep these limitations in mind. To underline that it is not true $g - r$ colour, we will denote this quantity as $(g - r)_\text{app}$.

\begin{figure*}
	\includegraphics[width=\textwidth]{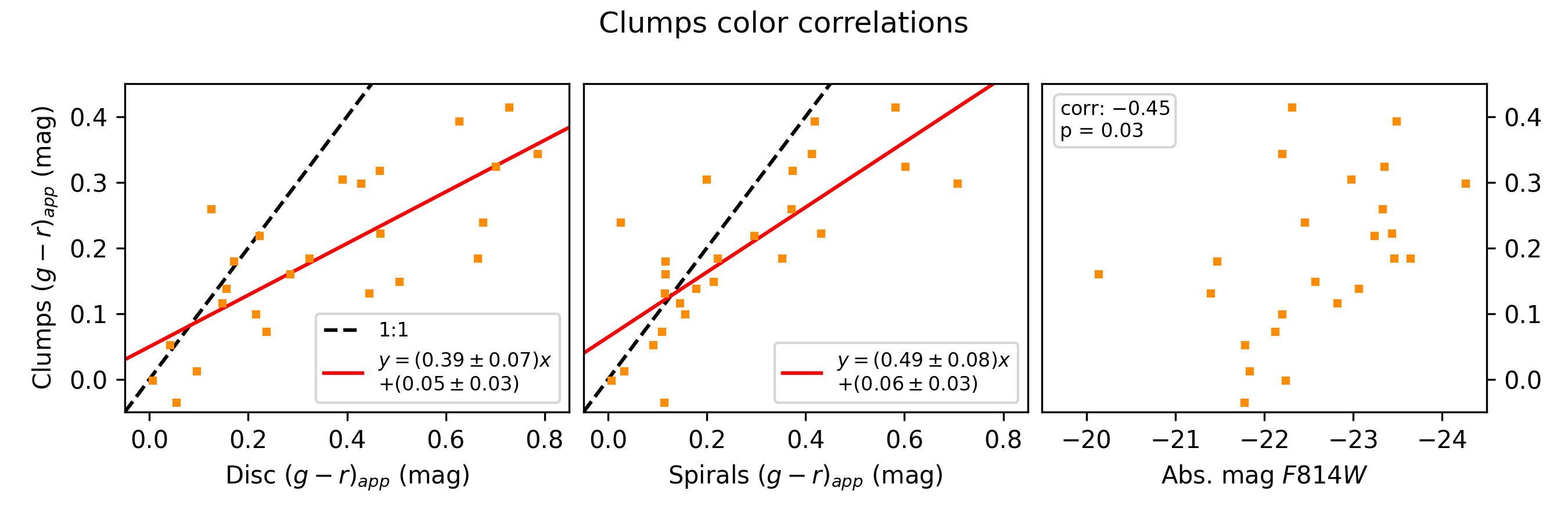}
	\caption{The relations between approximate colour $(g - r)_\text{app}$ of clumps and different parameters of a galaxy. Left: colour $(g - r)_\text{app}$ of the disc. Middle: colour $(g - r)_\text{app}$ of spiral structure. Right: absolute magnitude of a galaxy $M_\text{F814W}$. Dashed lines represent 1:1 relation. Solid lines show linear fits.}
	\label{fig:gr_comp}
\end{figure*}

We measure $(g - r)_\text{app}$ for the disc, spiral structure and clumps system in each galaxy: for two latter cases, we consider the combined flux from all spirals in a galaxy and same for all clumps. In Fig.~\ref{fig:gr_comp}, we compare $(g - r)_\text{app}$ for clumps, spirals and discs for individual galaxies. We see that clumps have bluer colours than both spirals and discs, but their colours correlate strongly. These correlations are not surprising, as both clumps and spirals are formed from the same disc parts. Although colour indices of clumps are closer to spirals than to discs, they are still different, most likely representing even higher specific star formation rate in clumps than in spirals. This difference also highlights that separating clumps and spirals is important. We note that clump colours correlate with galaxy luminosity which follows the fact that more massive galaxies tend to be redder than less massive ones \citep{Kajisawa2005, Santini2014}.

\subsubsection{SED fits}
\label{sec:sed_fits}

Compared to counting colour indices, SED fitting is a more complex tool for deriving physical properties of galaxies and their subsystems, and we applied it for 64 brightest clumps in our sample (see Section~\ref{sec:sed_fitting}). As MCMC analysis was also performed, it allowed us to measure not only parameters, but also a degree of their uncertainty. The most important parameters for us are clumps' masses and ages.

First, we note that in some cases we were unable to obtain a model SED that follows observations well. Often, the problem is the implausible photometry: fluxes can jump abruptly between filters where the continuum is expected to be smooth. This is probably caused by the fact that photometric models in \citelinktext{Chugunov2025a}{Paper I} were fitted independently in each band. They could converge to local minima representing slightly different structure, despite our efforts to control this. Even if the relative difference of models in adjacent bands is small, it could significantly impact residual images which are used for clumps identification. Thus, we select good SED fits (there are total 35 of them) as those where best-fit $\chi^2 < 2.5$, with error estimate being a sum of image noise at clump location and a 5\% fraction of flux. Visual inspection indicates that this threshold is useful to exclude unreasonable fit results.

\begin{figure}
	\includegraphics[width=\linewidth]{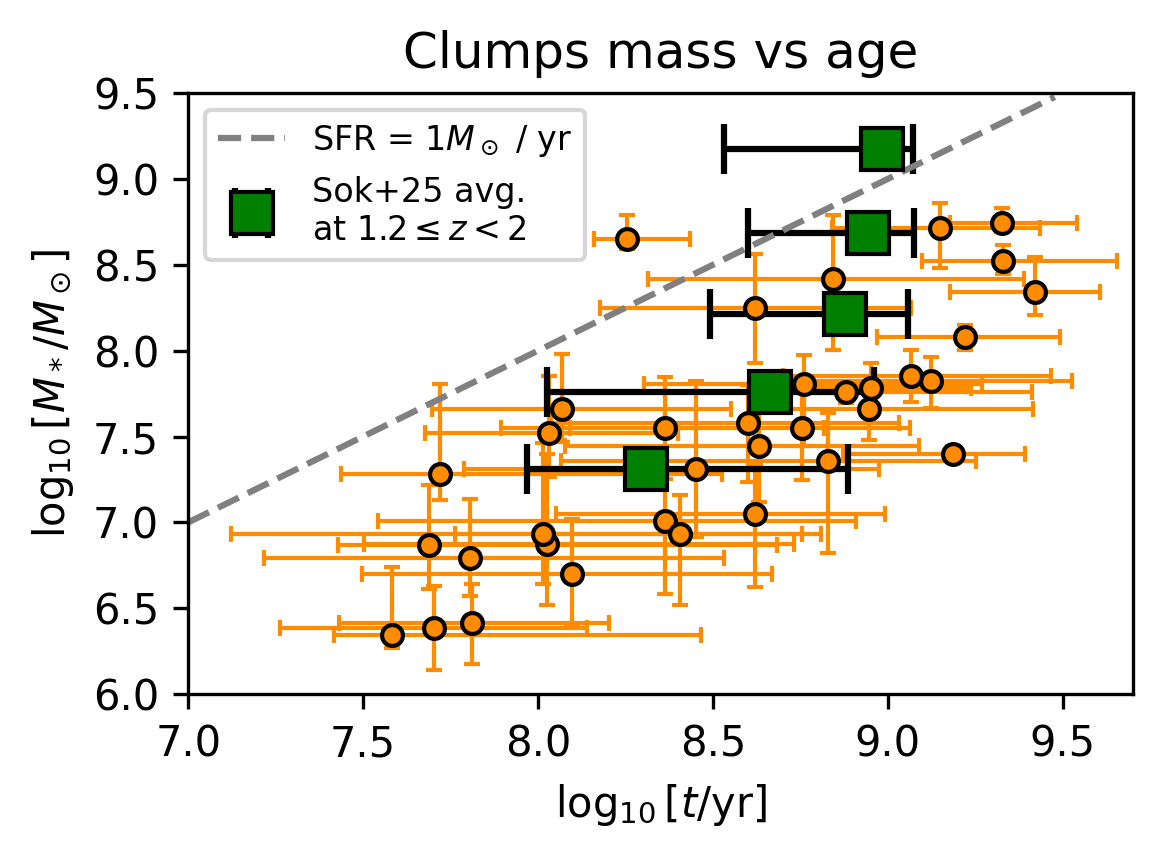}
	\caption{Clumps stellar mass versus age. For each clump, 1-$\sigma$ confidence interval is shown, based on MCMC analysis. For comparison, averaged ages in different mass bins from \citet{Sok2025} are shown. Dashed line shows possible mass evolution with time with constant SFR = 1 $M_{\sun}$ / yr.}
	\label{fig:mass_age}
\end{figure}

We compare measured ages and masses of clumps, as shown in Fig.~\ref{fig:mass_age}. There is a clear trend that older clumps tend to be more massive. The dependence between masses and ages of clumps was recently noted by \citet{Sok2025}, and our results align with their findings. They claim that this dependence may indicate that clumps' lifespan is defined by their mass, but we suggest that the same trend could be attributed to the observational effects: if old clumps with low mass exist, they would be faint and remain undetected. \citet{Claeyssens2023} also explain the same issue by the detection limits, noting the lack of old-age, low-mass clumps as well. Conversely, the apparent upper mass limit for a given age remains rather under-discussed, but it may point to the upper SFR limit of clumps. Finally, we note that mass-age diagram for clumps was constructed much earlier in \citet{Overzier2009}, but, despite the moderately visible trend, it was not discussed then.

\begin{figure}
	\includegraphics[width=\linewidth]{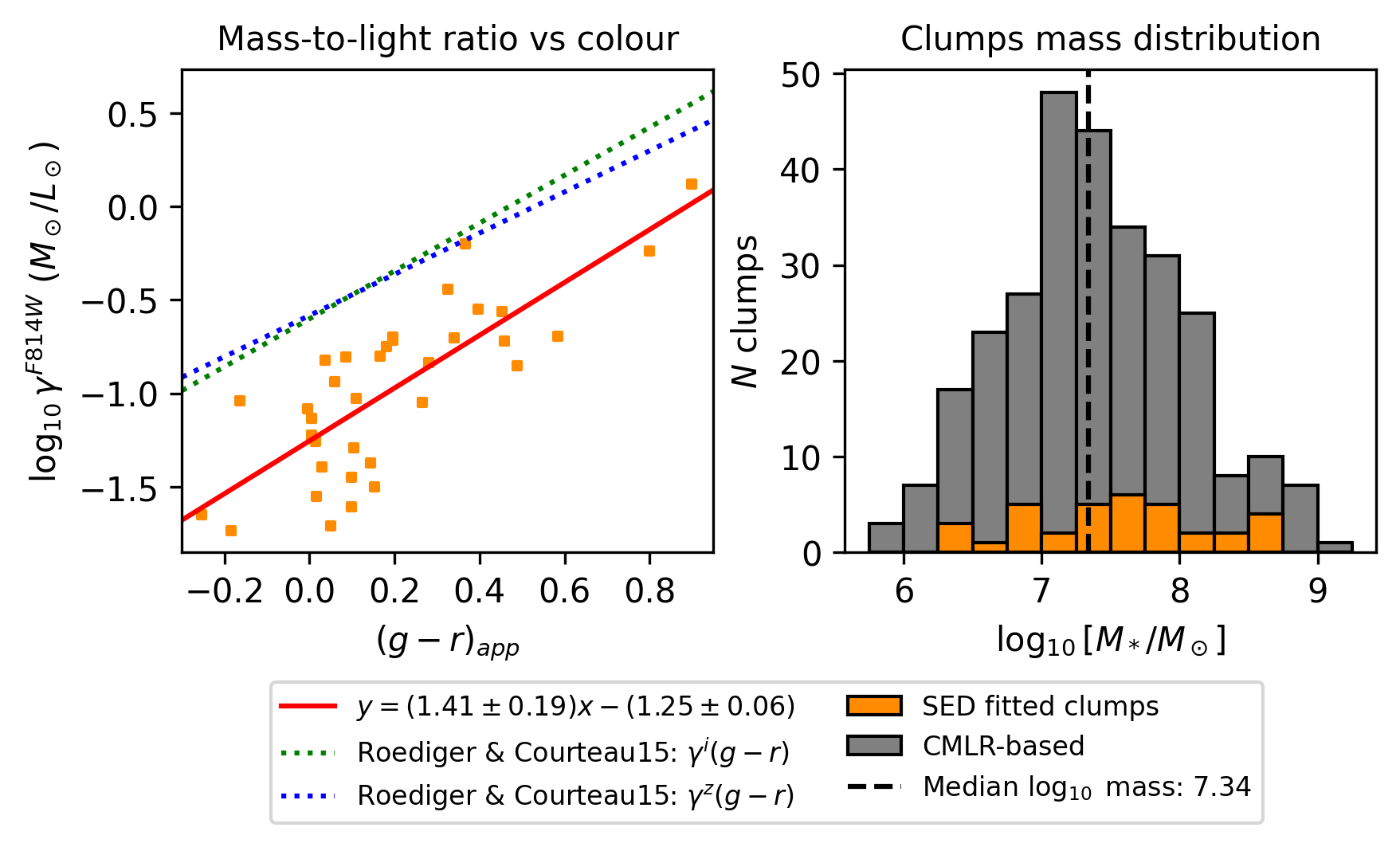}
	\caption{Left: diagram showing the relation between stellar mass-to-light ratio $\gamma^\text{F814W}$ and approximate colour index $(g - r)_\text{app}$ for clumps with good quality SED fitting. The linear fit (colour--stellar mass-to-light ratio, CMLR) is shown, as well as $i$ and $z$-band CMLR for local galaxies from \citet{Roediger2015}. Right: a stacked histogram showing the distribution of clumps by mass. For small number, it is determined from SED fitting, and for the majority of clumps, it is calculated from luminosity and colour via estimated CMLR.}
	\label{fig:mass_hist}
\end{figure}

As we have reliable measurements of mass and absolute magnitude of some clumps, we calculate their stellar mass-to-light ratio $\gamma^\text{F814W}$ and compare these values with approximate colour indices $(g - r)_\text{app}$. It turns out that there is a visible connection between these values, as shown in Fig.~\ref{fig:mass_hist}: redder clumps have higher mass-to-light ratio, and the median $\gamma^\text{F814W}$ is 0.1 $M_{\sun} / L_{\sun}$. The fact that such a dependence exists is not surprising at all: younger stellar populations have a fraction of massive stars which are over-luminous for their mass and blue. In fact, colour--stellar mass-to-light ratio relations (CMLR) are well established for local galaxies and commonly used \citep{McGaugh2014, Du2020}. The peculiarity in our results is that clumps $\gamma^\text{F814W}$ for a given colour is half order of magnitude smaller than commonly accepted values for local disc galaxies. In particular, we compare our results with CMLR of \citet{Roediger2015} in Fig.~\ref{fig:mass_hist} \citep[see also other CMLR in][]{Du2020}. However, this can be explained as galaxies, even if blue and dominated by young stars in radiation, still have a bulk of their mass in old stellar population. Conversely, clumps are expected to not have many old stars, so their mass-to-light ratio is naturally lower. It is important to mention, that \citet{Claeyssens2023} also provide masses and luminosities they measured for individual clumps, and their mass-to-light ratios, having a large scatter, are broadly consistent with ours. \citet{Telford2020} show theoretical mass-to-light ratios and colours dependent on the age of stellar population (see their Figure 5), which also agree with our values.

Furthermore, we can use the dependence of $\gamma^\text{F814W}$ on $(g - r)_\text{app}$ to estimate the masses without SED fitting. In Fig.~\ref{fig:mass_hist}, these mass estimates are shown together with SED fitted values. For clumps in our sample, the median stellar mass $\log M_*$ (in units of $\log_{10} [M_* / M_{\sun}]$) is nearly $7.3$ and the overall range is $6 \leq \log M_* \leq 9$. Both median and upper limit are smaller than commonly observed: for example, \citet{Kalita2025a} found the median $\log M_* = 9$ and the overall range $8 \leq \log M_* \leq 10$. \citet{Sok2025} results contain clumps with $6 \leq \log M \leq 10$, thus making their lower limit consistent with ours. \citet{Kalita2025b} reports $7.5 \leq \log M_* \leq 9.5$, with the average value of $8.5$; all deal with the redshift range similar to ours. Of course, studies invoking lensed galaxies report lower masses of clumps \citep[e.g.][]{Claeyssens2023, Claeyssens2025}. Nevertheless, the discrepancy of our results with more similar observational works has to be explained. As for \citet{Kalita2025a}, their algorithm probably disfavours the identification of faint clumps as they require the addition of each clump model to decrease Bayesian Information Criterion (BIC). Just as in the case of clump luminosities and sizes, the most likely explanations are either physical (clumps in spiral galaxies are intrinsically less massive than in clumpy ones), or methodological: the flux in spiral arms being already subtracted and deblending of clumps done (see Section~\ref{sec:CT}). However, there is a work by \citet{Zhu2026}: although they study clumps at $2 \lesssim z \lesssim 8$, they found median $\log M_* = 7.6$ and detect clumps with $\log M_* < 7$. Even though they explored much higher redshift range, the mass distribution at $z < 3$ looks roughly the same. We note that their algorithm of clumps also includes deblending, and they perform SED fitting with \verb|Prospector| as we do (see Section~\ref{sec:sed_fitting}).

\subsection{Clumps parameters and the proximity of spirals}

Are clumps in spiral arms and in inter-arm area are different? The answer to this question may point us to the physical nature of spiral structure. For example, if spirals are density waves, then one can expect a shock wave present in them, which may enhance star formation \citep[see theoretical considerations in][]{Roberts1969}. Recently, efforts were made to explore this effect in observations, for example, by comparing specific star formation rate in spiral arms and between them \citep[see][]{Querejeta2021, Querejeta2024, Sun2025}. As clumps are star-forming structures, they may be influenced by density waves as well, if they present in galaxies from our sample. Also, the local stability of the disc quantified by Toomre parameter \citep{Toomre1964} can be different in spirals and between them \citep[see e.g. Figure 1 in][]{Kostiuk2025b}, which represent different conditions for clump formation.

\begin{figure*}
	\includegraphics[width=\textwidth]{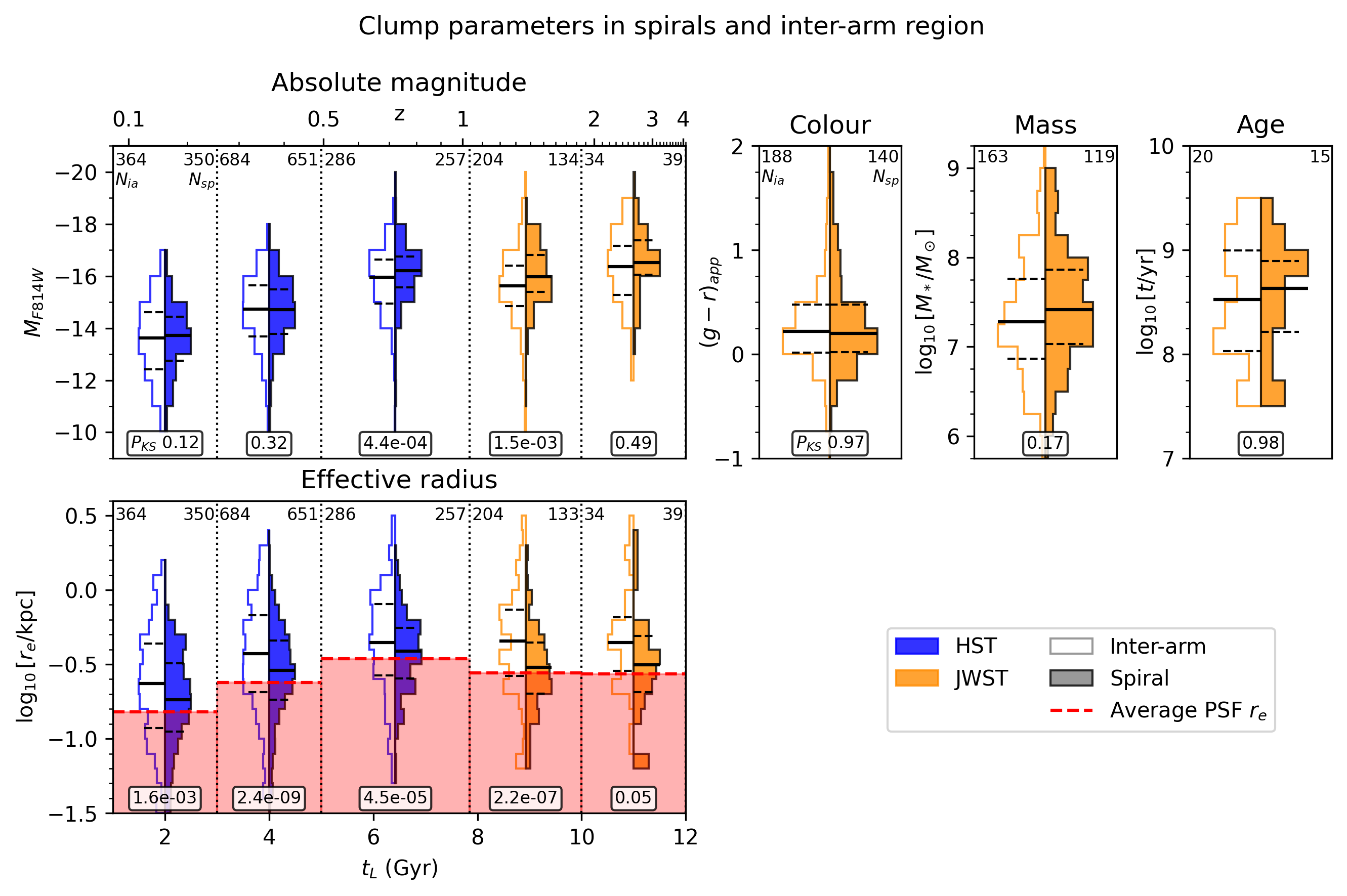}
	\caption{Histograms showing clumps distributions by different parameters, with clumps in spirals and clumps in inter-arm region are separated. For absolute magnitude $M_\text{F814W}$ (top left) and effective radius $r_e$ (bottom left), the data is binned by lookback time $t_L$. Colour $(g - r)_\text{app}$, mass $M_*$ and age $t$ (three boxes on top right) are not binned and only available for JWST subsample, at $z > 1$.\\
	Each histogram is two-sided, with left side (white) showing a distribution of inter-arm clumps, and right (coloured) representing clumps in spiral arms. Blue and orange colour represent HST and JWST subsamples, respectively. Solid horizontal lines in histograms represent median of a distribution, dashed ones show 25 and 75 percentiles. Vertical dotted lines mark bin borders. Numbers on top of each bin represent the number of clumps in this (left and right number stand for the number in inter-arm region $N_\text{ia}$ and the number in spirals $N_\text{sp}$). Numbers on the bottom are $p$-values of Smirnov test, which is a probability of that resulting samples of spiral and inter-arm clumps parameters being drawn from the same distribution.\\
	Only for effective radius, red dashed lines represent the relevant PSF $r_e$ at the given redshift of a given bin, and shaded region indicate that smaller $r_e$ are unreliable.}
	\label{fig:clump_comp}
\end{figure*}

As we gathered various clump parameters together and can differentiate clumps residing in spiral arms and in inter-arm region right now according to their visual positions (see Section~\ref{sec:location}), we can explore this matter further. In Fig.~\ref{fig:clump_comp}, we show the distributions of clumps by different parameters, for clumps in spirals and clumps in inter-arm region separately. We employ Smirnov two-sample test \citep[often confused with Kolmogorov --- Smirnov test, see][]{Berger2014} to judge whether the observed differences between two samples are statistically significant or not. For example, age is likely the most promising parameter to highlight differences in star formation, but the statistic is too small, as SED fitting was done for a limited part of the sample. However, colour indices which may also serve a rough proxy of star formation, were obtained for a larger sample, and also show no difference. For masses of clumps in spirals and inter-arm area, the difference is statistically insignificant, even if present. Absolute magnitudes and effective radii were measured for all clumps, and we compare them for spiral and inter-arm clumps in different redshift bins. Generally, we observe that clumps tend to be brighter and more compact in spirals than in inter-arm region.

First, the observed differences may be a manifestation of observational effects. For example, models of spiral arms from \citelinktext{Chugunov2025a}{Paper I} could fit large-scale clumps within spiral arms, and they did not appear in residual images there. Alternatively, the observed large objects (over 1--2 kpc in size) may be minor elements of spiral structure instead, which were not modelled in \citelinktext{Chugunov2025}{Paper I} and, as a result, they appear to be located outside modelled spiral arms. However, this is likely related only to a small number of outliers, as the difference of clump sizes is observed not only at extremely high $r_e$. The higher luminosity of clumps in spiral arms, in turn, may be caused by increased difficulty of identifying clumps in spiral structure, over less smooth background compared to inter-arm area.

On the other side, such a difference may have a physical reason. \citet{Inoue2018} developed a framework of spiral arm instability (SAI) for the process of spiral arms fragmentation and clump formation. SAI implies that, if the unstable perturbation is larger than width of spiral arm, then one can expect one-dimensional fragmentation and collapse of the spiral arm, which is different from classical axisymmetric Toomre analysis \citep{Toomre1964}. In fact, the most unstable wavelength in SAI is typically smaller than in axisymmetric case \citep[see Fig. 18 in][]{Inoue2018}, with the latter is likely applicable to inter-arm region. Also note that the predicted clump sizes in their work (1--2 kpc) are consistent with the largest structures we observe. Therefore, our results probably confirm that SAI mechanism has place, and may indicate that clumps are formed in existing spiral arms. And, because the gas density is increased in spiral arms, clumps there may be brighter compared to inter-arm region despite their smaller sizes. Note that even classical Toomre $Q$ parameter is a local quantity which can reflect different conditions across the disc, in particular in spiral arms and inter-arm region (see e.g. \citealt{Inoue2016}). Thus, \citet{Inoue2018} mention that different clump sizes in these parts of the disc can be expected in classical Toomre framework as well. In the same time, the lack of significant difference in colour and age is likely an evidence against any distinction between clumps in spiral arms and inter-arm region in terms of their specific star formation.

\section{Discussion}
\label{sec:discussion}

Clumps in distant galaxies are objects of interest, in particular due to their influence on the host galaxies, and there are plenty of recent works using state-of-art data from JWST and ALMA. However, high-$z$ spirals attract less attention as they are arguably more difficult to study. As a result, questions at the intersection, regarding the interplay between clumps and spirals, also remain unanswered. Our work approaches this area from observational perspective, using photometric data from HST and JWST, which reveals the stellar content of clumps.

Our study presents a sizeable sample of 3003 clumps in 159 galaxies. For comparison, \citet{Guo2018} presented 3193 clumps in 1270 galaxies. \citet{Sok2025} analysed 4500 galaxies and \citet{Sattari2023} studied 6767 galaxies which, as far as we know, is the largest sample of high-z galaxies searched for clumps. Modern works typically find that tens percent of high-$z$ galaxies are clumpy and usually detect one or a few clumps in a galaxy, so one can estimate that largest clump samples approach $\sim 10^4$ objects. Although the number of galaxies in our sample is modest, we find a relatively large number of clumps per galaxy (see Section~\ref{sec:results}), probably thanks to our algorithm involving photometric models with spiral arms and our sample consisting of well-resolved galaxies at moderate redshifts. However, this approach is not easily scalable as photometric decomposition with spiral arms is difficult to conduct. Nevertheless, it allowed us to achieve better sensitivity compared to more automated methods of clump identification and detect fainter clumps.

As we mentioned throughout the paper, in particular in Section~\ref{sec:CT}, the wide range of $z$ of galaxies in our sample makes their image quality significantly different. In principle, these observational effects could influence results in many aspects, from the incompleteness of galaxy sample to the detection limits of clumps; this is a common class of problems in studies of high-$z$ galaxies and a lot of attention is usually paid to mitigate it, see for example \citet{Kalita2025a, Sok2025}. However, our work specifically deals with well-resolved spiral galaxies and, which is the most important, our goals were neither to perform a census of clumps down to the faintest ones, nor to precisely measure the evolution of clumps over the cosmic time. Instead, we mostly aimed to compare the relative properties of clumps and spiral structure. When applicable, we consider the results in limited redshift bins or check how observed properties depend on SNR, ensuring observational effects do not influence our conclusions.

Subtraction or other possible way to account for spiral arms is also important for another reason. If spiral arms are neglected, then the flux of spiral arms will be attributed to the clumps. As our work, to the best of our knowledge, is the first to deal with spiral arms specifically, this may explain that our results regarding clumps parameters differ from the literature. In our study, clumps turn out to be smaller, less massive, and their contribution to the total luminosity of a galaxy is lower, both individually and combined.

However, our results do not suggest changes to the established fundamentals of clumps nature in general. For example, we confirm clump-to-total ratio varying with wavelength, increasing towards shorter wavelengths \citep{Messa2022}. The fact that clumpiness with time peaks at $z \sim 1\ldots2$ is also in agreement with the literature \citep{Murata2014, Shibuya2016, Sattari2023, delaVega2025}. Same holds for the correlation between luminosity and size \citep{Cava2018, Claeyssens2023, Kalita2025a}, and for anti-correlation between clump-to-total ratio and bulge-to-total ratio \citep{Kalita2024}. Age-mass dependence does not contradict earlier results as well \citep{Sok2025}. All these comparisons with the literature on the subject may serve as some sort of validation of our results.

On the other side, our sample of galaxies contains exclusively spirals unlike in most other studies devoted to clumps. Therefore, it is simply not fully correct to compare clumps in general population and exclusively spirals. Alternative explanation for our discrepancy with other results could be that clumps in spiral galaxies are possibly different from ones in clumpy galaxies. For now, we cannot point to the ultimate reason of the mentioned differences of our results and literature. To find which of these reasons (or possibly both of them) influence different properties, it would be helpful to conduct a study of clumps in both spiral and clumpy galaxies in a consistent way, comparing these structures between two types of galaxies. The work of \citet{Elmegreen2009} is a good example of a study paying attention to this matter, and modern data from JWST and ALMA could be used to update their findings.

Another uncommon aspect of our work is that we do not only explore the properties of clumps, but also compare them directly to the characteristics of spiral arms. In particular, we examine the degree and character of spatial distribution of clumps and their concentration towards spirals, check the overall connection between the properties of spirals and clumps (such as the contribution to the total luminosity, pitch angle and asymmetry). Finally, we check if clumps are different in spiral arms and between them, or not. Summarising the results on this matter together, we can conclude that clumps are likely connected to spirals. This is illustrated by the observed concentration of clumps towards spirals and the correlation between some properties of clumps and spirals. Clumps in spiral arms tend to be smaller but brighter than in inter-arm area, and there is no difference in colours and ages of them. Combined, these properties may indicate that spiral arms enhance clumps formation, in a way similar to what was shown by \citet{Kim2002} for spurs formation. More specifically, the difference in size indicates that clumps in spiral arms are possibly formed as a result of spiral arm instability \citep{Inoue2018}, being an evidence in favour of this mechanism. And the lack of difference in colours and ages of clumps occupying different parts of the disc may indicate that star formation characteristics of these objects (such as specific star formation rate) are not changed by spiral arms. In other words, clumps formation seems to be gathered in spirals arms, but does not have increased star formation efficiency and modified properties of individual star formation regions. Note that recent studies of local galaxies focused directly on the latter question, tend to support the same conclusion \citep[e.g.][]{Querejeta2021, Querejeta2024, Sun2025}, although in some conditions star formation seem to be boosted rather than gathered in spirals \citep{Yu2021}. We also show that clumps and spirals have different photometric properties (e.g. they differ in colour) which also highlight the importance of their separate modelling. However, our findings are still not enough to decisively support one or another theory of spiral formation. For example, observed properties of clumps can be explained in both density wave theory and in dynamical spirals framework \citep{Sellwood2022a}, because in both cases spiral arms can act as sites of large-scale gas concentration.

There is still a room for improvement of our methods. It is reasonable to assume that structural properties of a galaxy vary smoothly with wavelength; as a consequence, photometric model parameters should not change abruptly between adjacent filters. In practice, when multiwavelength decomposition is done, the models are often prepared individually in each filter \citep[see examples of][]{Rios-Lopez2021, Gong2023, Marchuk2024b, Marchuk2025, Chugunov2025a}. However, this may lead to inconsistencies between different filters which are very difficult to control; in our current work, they became problematic for SED fitting in particular (see Section~\ref{sec:sed_fits}). The much more robust and effective approach is a simultaneous fitting of images in multiple filters, where model parameters are not fully independent between them. It was implemented in \verb|MegaMorph| project \citep{Haussler2013, Vika2013} as an extension of \verb|GALFIT| fitting package \citep{Peng2002}, called \verb|GALFITM|. Photometric modelling with \verb|GALFITM| is relatively common, although in most cases it is limited to single S{\'e}rsic fitting \citep{Vulcani2014, Kennedy2015, Mosenkov2019}. On the other side, there are also examples of 2- and 3-component multi-band fitting with \verb|GALFITM| \citep{Kruk2018, Buzzo2021}. However, adding custom functions to \verb|GALFIT| is no straightforward task, which is critical for our study as we make extensive use of models of spiral arms. Thus, the viable option for the future studies is to create an extension for \verb|IMFIT| \citep{Erwin2015} to implement simultaneous multiwavelength fitting.

The overall design of our work also may be improved. Most importantly, the current work relies heavily on our \citelinktext{Chugunov2025a}{Paper I}, as we use photometric models of galaxies with spiral arms. However, our previous work originally was not intended to have any connection to clumps, or to have a following study in this direction. As a result, the main downsides of our approach are that our sample lacks galaxies without spiral structure and even for spiral galaxies we rejected some filters where spiral structure looks to patchy. In fact, this often resulted in rejecting images blueward from the Balmer break, which ultimately limited the capabilities of SED fitting. Also, there was no consistent approach to clumps in \citelinktext{Chugunov2025a}{Paper I} (like universal masking), and modelling of clumps and spirals can be made more coordinated. All these possible improvements may be done in future studies.

Broadly speaking, one may consider the inclusion of clumps (or star-forming regions in general) in decomposition as the logical extension of modelling spirals, although is was implemented in this work for the first time. Clumpy spirals are the natural domain to use this approach. Another similar example is ultraviolet images of local galaxies where spirals become extremely patchy, and the smooth part of light distribution in spirals is almost lost beneath individual star-forming regions \citep[as observed in examples of][]{Marchuk2024b, Kostiuk2025a, Chugunov2025b}.

In any case, the simultaneous modelling and studying spiral arms and clumps may shed light on the question of how star formation in the Universe changed from the ``clumpy'' regime to more ``ordered'' regime. Nowadays, there are a few competing theories of spiral structure formation \citep[see a review by][]{Sellwood2022a}. Paying attention to clumps or star-forming regions in the context of spiral arms may also help answer the ultimate question of the nature of spiral arms. Particularly regarding our results on clumps, they provide some insights on the spiral structure. For example, the increased asymmetry of the spiral structure at high-$z$ domain seems to appear independently of clumps and should have a different explanation, opposite to our assumption in \citelinktext{Chugunov2025a}{Paper I}. For example, galaxies in the early Universe had a higher interaction rate \citep{Lavery2004, Conselice2022} which can lead to that spiral structures within discs are more often perturbed. Notably, the fraction of warped discs is also increased at high $z$, which also independently points to the role of tidal interactions \citep{Reshetnikov2025}. The gradual shift of the dominating mechanism of spiral formation is also a possible explanation. In \citelinktext{Chugunov2025a}{Paper I} and \citet{Reshetnikov2022, Reshetnikov2023}, we have brought some evidence that the dominant mechanism may change.

For our study, we selected well-resolved spiral galaxies. This have led us to that our objects have lower redshifts than in many modern studies, and a large part of galaxies in our sample has $z < 0.5$. Meanwhile, clumpy galaxies presumably being gradually replaced by spirals over this transitional redshift range, which makes it important. For example, simulations done by \citet{Renaud2024} indicate that clumps at cosmic noon follow the same scaling relations as local star-forming clouds, probably suggesting a continuous transition in parameters. Indeed, low-redshift clumps in our sample are mostly smaller and fainter than ones near cosmic noon (although observational effects should be kept in mind), thus being closer to the local star-forming regions.

Although SED fitting is a powerful tool of measuring the physical properties of galaxies and their components such as clumps, its potential is limited in our work. In most cases, filters cover only rest-frame optical and near-infrared wavelengths. This is not enough to break the degeneracy between the age, metallicity and dust attenuation \citep[for details, see][]{Conroy2013, Pacifici2023}. The increase of each of them will lead to reddening, and, for a given SED, it is impossible to discern the old stellar population and young but dusty one, using only broad-band filters in optical and near-infrared. In fact, this is the reason why we applied MCMC analysis to SED fitting: we aimed to estimate how (un)reliably we measure these values, given the small amount of information we have. As expected, the uncertainty in age is large, sometimes exceeding an order of magnitude. However, we still observe at least a fraction of massive clumps to have ages $>10^8$ yr and even exceeding $10^9$ yr. This result indicates that some of clumps are long-lived, contributing to the ongoing debate over this question \citep{Kalita2025a}.

One potential option to resolve these problems is spectroscopy, but unfortunately it is extremely time-consuming for high-$z$ galaxies. Aside from that, one could break the degeneracies and avoid catastrophical failures by extending wavelength range to either UV (which traces the current star formation) or to mid-IR (associated with the interstellar medium radiation), or to both \citep{Battisti2025}. Unfortunately, the radiation in wavelengths blueward from the Balmer break is much fainter than in optical, and it becomes harder to explore the fine details of galaxy structure there. Regarding mid-IR domain, spatial resolution becomes the main problem here, as it decreases with increasing wavelength for a given aperture. Although JWST has MIRI instrument for longer wavelengths than NIRCam (up to 25 $\mu$m), its resolution is not good enough to resolve clumps at $z > 1$.

On the other side, Atacama Large (sub-)Millimeter Array may be immensely useful for this matter. As an instrument which resolution matches HST and JWST, it allows one to map dust and gas content of high-$z$ galaxies and to trace gas kinematics \citep[see a review by][]{Hodge2020}. Various studies explored clumps in high-$z$ galaxies with ALMA \citep[e.g.][]{Carniani2017, Cibinel2017}. Some recent studies already made use of combining observations from ALMA and JWST, even though ALMA data often has resolution not as good as possible \citep{Kalita2025a}, or may be limited to a single object \citep{Fujimoto2025}. It would be beneficial to have a sample of galaxies with combined and consistent high-resolution observations by HST, JWST and ALMA. This would provide a multiwavelength coverage from rest-frame UV to far-IR, ideally in the same manner as the DustPedia project for the local Universe \citep{Clark2018}.

\section{Conclusions}
\label{sec:conclusions}

In this Section, we summarise the most important findings of our study. In this work, we have performed photometric analysis of clumps in 159 spiral galaxies at $0.1 \leq z \leq 3.3$, observed by HST and JWST, finding 3003 clumps in overall. For the first time, such analysis was done with spiral structure modelled separately from clumps in \citelinktext{Chugunov2025a}{Paper I}. For a fraction of our sample, we applied SED fitting to obtain the parameters of clumps' stellar populations.

\begin{enumerate}
	\item We have confirmed previous results regarding clumps-to-total dependence on wavelength: clumps-to-total ratio is higher at shorter wavelengths (see Fig.~\ref{fig:CT_wavelength}). We confirm that clumpiness evolves with time, with a peak at $z \sim 1 \ldots 2$ (see Figs.~\ref{fig:CT_true_z},~\ref{fig:CT_three_pars}).
	\item We found clump-to-total ratio to correlate negatively with bulge-to-total ratio and positively with spiral-to-total ratio, indicating a connection between clumps and spirals. There is no significant correlation with the asymmetry of spiral structure, meaning that the asymmetry evolution with time (see \citelinktext{Chugunov2025a}{Paper I}) is probably not caused by evolving clumpiness of galaxies (see Figs.~\ref{fig:CT_params}).
	\item We found clumps to be concentrated towards spiral arms: 20\% pixels of a galaxy with the highest brightness of spirals contain 47\% of all clumps, on average. This result is caused by the coincidence in both radial and azimuthal distribution of clumps and spirals across the host galaxy (see Figs.~\ref{fig:rch},~\ref{fig:clump_loc_hist}). The degree of concentration of clumps towards spirals correlates weakly positively with spiral-to-total ratio and bulge-to-total ratio, and negatively with clump-to-total ratio and spiral asymmetry. This can be interpreted as the increase of clumps concentration towards spirals represents the ``ordering'' or star formation across the galaxy (see Fig.~\ref{fig:fcs_params}).
	\item We found clumps' mass and age to be connected, with older clumps being more massive. We found a color--stellar mass-to-light ratio (CMLR) for clumps and observe a median mass-to-light in rest-frame F814W filter to be 0.1 $M_{\sun} / L_{\sun}$ (see Figs.~\ref{fig:mass_age},~\ref{fig:mass_hist}).
	\item We examined the difference in clumps parameters in spiral arms and in inter-arm area. We found no difference in age and colour, possibly supporting that spirals function as gatherers of star formation, and found that clumps are smaller but brighter in spirals compared to inter-arm area, which supports the scenario where clumps in spirals are formed via spiral arm instability (SAI) \citep{Inoue2018} in existing spiral structures (see Fig.~\ref{fig:clump_comp}).
	\item We found clumps to be generally smaller, less massive and contribute less to the luminosity compared to the literature. This is most likely caused by the fact that we account for spiral structure in our photometric models, or because clumps are physically different in spiral galaxies, compared to clumpy ones (see Sections~\ref{sec:CT},~\ref{sec:sizes} and~\ref{sec:sed_fits}).
\end{enumerate}

Overall, our results emphasize the connection of clumps and spiral arms at high redshifts, but also the differences between these two spatial modes of star formation in galaxies. Our findings highlight the importance of analysing these structures separately from each other for correct retrieval of their parameters, and demonstrate the usefulness of this approach to explore clumps and spirals, as well as interplay between them. In our work, we attempted to bridge a gap between studies of local spiral galaxies where star-forming regions are small and secondary compared to the ordered spiral structure, and high-redshift clumpy galaxies. We hope our findings can be helpful for discovering how the transition from one regime to another occurred in the Universe.

\section*{Acknowledgements}

This work is based in part on observations made with the NASA/ESA Hubble Space Telescope obtained from the Space Telescope Science Institute, which is operated by the Association of Universities for Research in Astronomy, Inc., under NASA contract NAS 5–26555, for the COSMOS survey.

This work is based in part on observations made with the NASA/ESA/CSA James Webb Space Telescope. The data were obtained from the Mikulski Archive for Space Telescopes at the Space Telescope Science Institute, which is operated by the Association of Universities for Research in Astronomy, Inc., under NASA contract NAS 5-03127 for JWST, for CEERS and JADES surveys.

We thank the anonymous reviewer for their comments and suggestions that helped us to improve the paper.

\section*{Data Availability}

Images, photometric models and tables of parameters are available at \url{https://github.com/IVChugunov/Distant_spirals_decomposition}. All other data underlying this article will be shared on reasonable request to the corresponding author.



\bibliographystyle{mnras}
\bibliography{main}


\bsp	
\label{lastpage}
\end{document}